\documentclass[11pt]{article}

\usepackage[a4paper,margin=1in]{geometry}
\usepackage{amsmath,amssymb,amsfonts,bm}
\usepackage{graphicx}
\usepackage{mathtools}
\usepackage{array}
\usepackage{longtable}
\usepackage[hidelinks]{hyperref}

\newcommand{\Rmem}{R_{\mathrm{ves}}}                 
\def\endens{\varpi}
\newcommand{\Order}[1]{\mathcal{O}(#1)}
\newcommand{\dd}{\mathrm{d}}
\newcommand{\eps}{\epsilon}
\newcommand{\Hb}{\tilde{H}_0}          
\newcommand{\thetac}{\theta_c}
\newcommand{\Jcat}{\mathcal{J}}        
\newcommand{\II}{\mathrm{I\!I}}
\newcommand{\meanc}{\mathcal H}        
\newcommand{\Seff}{\Sigma_{\mathrm{eff}}}       

\title{\bfseries Boundary-layer analysis of the partial engulfment of a small particle by a lipid membrane}

\author{Gaetano\ Napoli\thanks{E-mail: \texttt{gaetano.napoli@unina.it}}\\[2pt]
{\small Dipartimento di Matematica e Applicazioni ``Renato Caccioppoli'',}\\
{\small Universit\`a degli Studi di Napoli ``Federico II'', Napoli, Italy}}

\date{\today}

\begin{document}
\maketitle
\begin{abstract}
\noindent
We study the axisymmetric partial engulfment of a small rigid sphere by a fluid Helfrich membrane. The neck that joins the wrapped cap to the membrane is an elastic boundary layer, which we analyse by matched asymptotic expansions. Its leading surface is a catenoid, which stores no bending energy and is obtained from the equilibrium equations for a regular inner expansion. Beyond the term linear in the size ratio, known from the literature, the energy contains a term proportional to the logarithm of the size ratio. Its coefficient is the effective lateral tension acting on the logarithmic excess area of the catenoid, and it depends on how the membrane is closed far from the particle. For a membrane in contact with a tension reservoir we obtain the wrapping energy to second order in the size ratio, in closed form at every wrapping angle. The resulting landscape contains as limits the low-tension regime and the local landscape of the wrapping problem previously derived in the literature, and describes the crossover between them. The landscape recovers the known binding threshold of a flat membrane, independent of tension and spontaneous curvature, and the bistability of the partially and fully wrapped states is controlled mainly by the spontaneous curvature. The order of the remainder is estimated, and the neck energy is checked against numerical solutions of the full nonlinear shape equations.
\end{abstract}

\medskip
\noindent\textbf{Keywords:} Helfrich membrane; matched asymptotics;
elastic boundary layer; catenoid; Jacobi operator; minimal surfaces;
hysteresis; tension reservoir; particle wrapping.

\section{Introduction}\label{sec:intro}
The adhesion of small rigid inclusions to lipid membranes, and their partial or complete engulfment, underlie the passive, elastic component of processes such as endocytosis, viral budding and the cellular uptake of engineered nanoparticles~\cite{Lipowsky1991,Seifert1997,ACL2015}. In the continuum description, a rigid sphere of radius $a$ gains adhesion energy over the region on which it is wrapped, and pays bending energy both on the wrapped cap and on the neck that connects the cap to the surrounding membrane~\cite{DesernoBickel2003,Deserno2004,NapoliGoriely2020,LipowskyDoebereiner1998,ACL2015,ACL2016,ACL2017}. When the particle is much smaller than the membrane, the size ratio $\eps=a/\Rmem$, with $\Rmem$ the macroscopic radius of the membrane, is small. Consequently, the cap and the neck occupy a region of size $\Order{a}$ around the contact line, whereas the rest of the membrane varies on the macroscopic scale $\Rmem$. The neck is then an elastic boundary layer~\cite{BiscariNapoli2007}, that is, a thin region of rapid geometric variation that joins two slowly varying outer states; its shape can be determined by matched asymptotic expansions even when the outer shape is known only implicitly.

Boundary-layer treatments of wrapping usually proceed in two stages. First, the inner problem is solved at leading order, which yields the familiar catenoidal neck~\cite{DesernoBickel2003,Deserno2004}; second, the free energy is evaluated on that surface. This procedure meets a difficulty that is easily overlooked. The leading inner surface is minimal, and the Helfrich density is a function of the mean curvature measured from the spontaneous curvature, which is of higher order in the size ratio; hence the density vanishes pointwise, and the catenoidal neck contributes no energy at leading order. This holds at leading order only. The energy that the inclusion adds to the membrane through the neck is therefore not a bending energy of the leading surface but an excess energy of the boundary layer, which appears at the next orders. Its determination is the aim of this paper.

\emph{Relation to previous work.} Several results used below are not new. The threshold for the onset of adhesion and its independence of the tension are classical~\cite{LipowskyDoebereiner1998,DesernoBickel2003,Deserno2004}. Agudo-Canalejo and Lipowsky derived the stability limits of the free and of the completely engulfed state~\cite{ACL2015}, the expansion near complete wrapping for a narrow neck~\cite{ACL2016}, and the landscape of a small particle on a membrane of arbitrary shape~\cite{ACL2017}. In Refs.~\cite{ACL2015,ACL2017} the completely engulfed state is a limit shape with an ideal neck that costs no energy. In Ref.~\cite{ACL2016} a narrow neck is represented by a trial contour of segments of constant mean curvature, which does not satisfy the shape equations, and the resulting expansion in the neck radius is validated by numerical energy minimization. For a flat membrane under tension and without spontaneous curvature, Deserno~\cite{Deserno2004} obtains analytically the energy of the free membrane in the small-gradient approximation, asymptotically exact for small penetration; at finite wrapping angle the exact energy is computed numerically. Wrapping driven by an active or external force, which has a nonzero axial force, is not treated here~\cite{Xiao2022}.

The present work makes three contributions. First, it derives the catenoidal neck from the shape equations, as the leading term of a regular inner expansion and not as an assumption; the higher orders of the expansion then bound the energy terms that the catenoid leaves out. Second, it determines the next term of the neck energy, of order $\eps^{2}\ln(1/\eps)$, which the descriptions of Refs.~\cite{ACL2015,ACL2016,ACL2017} do not contain. For a tension reservoir it is obtained in closed form at every wrapping angle and also for $c_0\neq0$; for $c_0=0$ and small penetration it reduces to the result of Ref.~\cite{Deserno2004}, whose energies at finite wrapping angle are computed numerically. Third, the resulting landscape contains as limits the description of Deserno, for vanishing spontaneous curvature, and the local landscape of Agudo-Canalejo and Lipowsky, when the logarithmic term is negligible, and describes the crossover between them (\S\ref{sub:hysteresis}).

The inner problem, at given contact data, is that of the conical inclusion of Biscari and Napoli~\cite{BiscariNapoli2007}, a rigid cone of prescribed apex angle in an axisymmetric vesicle of arbitrary outer shape, for which they obtained the second-order correction of the neck in closed form. Here, however, the contact angle is a free variable, determined by the equilibrium at the contact line; the energy of the neck enters the landscape of the wrapped state; and the tension and the closure of the outer field fix the coefficient of the logarithm.

\emph{Scope.} The logarithmic term does not control the landscape in general: it is smaller than the term linear in the size ratio by a factor of the order of the size ratio times its logarithm, and it dominates only for a nearly flat membrane under tension and with negligible spontaneous curvature, where it is the tension acting on the logarithmic excess area of the catenoid (the regime of the tension-induced barrier of Ref.~\cite{Deserno2004}, \S\ref{sub:tense}). For a closed vesicle the leading-order landscape is the local one of Ref.~\cite{ACL2017}. Beyond the $\Order{\eps}$ slope-flux term of that reference (\S\ref{sub:landscape7}), the bending energy of the layer is quadratic in the $\Order{\eps}$ correction to the catenoid, the logarithm is the tension acting on the excess area of the catenoid, and both depend on the outer closure. We carry out this closure for a tension reservoir and, for a closed vesicle, when the mean curvature of the outer membrane at the particle equals the spontaneous curvature (\S\ref{sub:equalcurv}); the general closed vesicle is left to future work.

\section{Equilibrium equations}\label{sec:model}
\begin{figure}[t]
\centering
\includegraphics[width=0.42\textwidth]{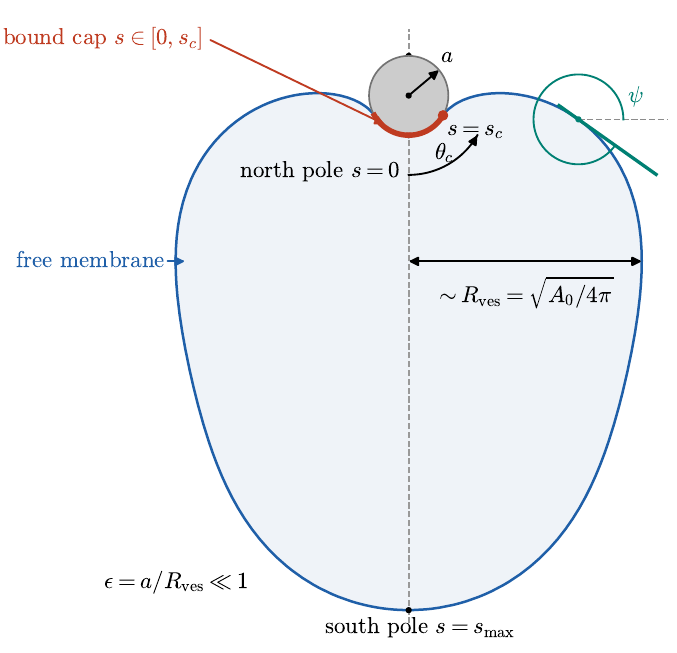}
\caption{Axisymmetric geometry of the model. A rigid sphere of radius $a$
(grey) adheres at the north pole ($s=0$) of a fluid membrane of area $A_0$; the
meridian is parametrized by the arclength $s$ and the tangent angle $\psi$, measured from the horizontal. The
membrane is \emph{bound} to the particle over the cap $s\in[0,s_c]$ (red),
where it conforms to the sphere, and \emph{free} beyond the contact line
$s=s_c$, at wrapping angle $\thetac=s_c/a$, up to the outer end $s=s_{\max}$ (the south pole in the drawing; in the reservoir problem $s_{\max}\to\infty$, the far field is flat and no closure at a far pole is used). The macroscopic radius is $\Rmem\equiv
\sqrt{A_0/4\pi}$, and the size ratio is $\eps=a/\Rmem\ll1$. The neck that joins the bound cap to the free
membrane is the elastic boundary layer analysed in \S\ref{sec:asymptotics}.}
\label{fig:schematic}
\end{figure}

We consider a rigid spherical particle of radius $a$ that adheres, with adhesion energy per unit area $\Delta\gamma>0$, to a fluid membrane governed by the Helfrich energy~\cite{Canham1970,Helfrich1973}
\begin{equation}
W_{H}=\frac{k}{2}\int_{\mathcal S}\big(2H-2c_0\big)^{2}\,\dd A .
\label{eq:helfrich}
\end{equation}
Here $k$ is the bending rigidity, $H=(c_m+c_p)/2$ the mean curvature, with $c_m$ and $c_p$ the principal curvatures, and $c_0$ the spontaneous curvature, that is, the preferred value of $H$; the integral extends over the whole membrane surface $\mathcal S$. Only axisymmetric configurations are considered (Fig.~\ref{fig:schematic}).

The meridian $(\rho(s),z(s))$ is parametrized by the arclength $s$, measured from the north pole, and by the tangent angle $\psi(s)$, so that
\begin{equation}
\rho'=\cos\psi,\qquad z'=\sin\psi .
\label{eq:frame}
\end{equation}
The mean curvature is then
\begin{equation}
2H=-\Big(\psi'+\frac{\sin\psi}{\rho}\Big),
\label{eq:H}
\end{equation}
where $'=\dd/\dd s$ and the sign of $2H$ refers to the outward normal. The principal curvatures are $c_m=-\psi'$ (meridional) and $c_p=-\sin\psi/\rho$ (parallel). The membrane is attached to the particle at the north pole and conforms to it over the bound region $s\in[0,s_c]$, where $\psi=s/a$, $\rho=a\sin(s/a)$ and $H=-1/a$. The contact line is at $s=s_c$; it defines the wrapping angle $\thetac=s_c/a$.

Consequently the bending energy~\eqref{eq:helfrich} splits over the bound cap $\mathcal S_{\mathrm b}$ and the free membrane $\mathcal S_{\mathrm f}$, with $\mathcal S=\mathcal S_{\mathrm b}\cup\mathcal S_{\mathrm f}$,
\begin{equation}
W_{H}=\underbrace{\frac{k}{2}\!\int_{\mathcal S_{\mathrm b}}\!(2H-2c_0)^{2}\dd A}
_{\textstyle W_{\mathrm b}=4\pi k(1+a c_0)^{2}(1-\cos\thetac)}
\;+\;\frac{k}{2}\!\int_{\mathcal S_{\mathrm f}}\!(2H-2c_0)^{2}\dd A .
\label{eq:Ebend_split}
\end{equation}
The bound contribution is an explicit function of the wrapping angle, because $2H=-2/a$ on the cap. The adhesion of the bound region lowers the energy in proportion to the contact area,
\begin{equation}
W_{\mathrm{ad}}=-\Delta\gamma\,A_{\mathrm b}
=-2\pi a^{2}\,\Delta\gamma\,(1-\cos\thetac),
\label{eq:adhesionenergy}
\end{equation}
where $A_{\mathrm b}=2\pi a^{2}(1-\cos\thetac)$ is the bound area, so that the total energy is $W=W_{H}+W_{\mathrm{ad}}$.

Adhesion is measured by the dimensionless gluing number
\[
\eta=a\sqrt{\Delta\gamma/k},
\]
and the particle size by
\begin{equation}
\eps=\frac{a}{\Rmem}\ll1,\qquad \Rmem=\sqrt{A_0/4\pi},
\label{eq:eps}
\end{equation}
where $A_0$ is the total membrane area. In the reservoir problem $A_0$ plays no other role: the physical small parameters are $ac_0$ and $\Seff a^{2}/k$, with $\Seff$ the effective tension defined below, and $\eps$ is a bookkeeping parameter that orders them (\S\ref{sec:reservoir}, \S\ref{sub:hysteresis}).

The equilibrium meridian of the free membrane extremizes $W=W_{H}+W_{\mathrm{ad}}$ under the geometric relations~\eqref{eq:frame} and under the area constraint, which is enforced by the membrane tension $\Sigma$ acting as a Lagrange multiplier. The membrane is in contact with a reservoir of area at zero excess pressure, so that the osmotic pressure vanishes, $P=0$, and no volume constraint arises; the general closed vesicle, with its two global constraints, is not treated here (a special case is considered in \S\ref{sub:equalcurv}). The augmented functional is
\begin{equation}
\mathcal{F}=W_{H}+W_{\mathrm{ad}}+\Sigma A,\qquad A=A_{\mathrm b}+A_{\mathrm f},
\label{eq:augmented}
\end{equation}
where $A_{\mathrm f}=2\pi\!\int_{s_c}^{s_{\max}}\!\rho\,\dd s$ is the area of the free membrane. The coordinate $s\in[0,s_{\max}]$ runs from the north pole to the outer end $s_{\max}$, which lies at infinity in the reservoir.

The bound cap is rigid, its shape being that of the sphere, so that only the free portion $s\in[s_c,s_{\max}]$ is varied. The geometric relations~\eqref{eq:frame} are enforced pointwise as well: $\rho$, $z$ and $\psi$ are treated as independent fields, and two multiplier fields, $\lambda_\rho(s)$ and $\lambda_z(s)$, are introduced. The reduced Lagrangian density (per $2\pi$) is
\begin{equation}
\endens=\frac{k}{2}\,\rho\,(2H-2c_0)^{2}+\Sigma\,\rho
+\lambda_\rho\,(\rho'-\cos\psi)
+\lambda_z\,(z'-\sin\psi),
\label{eq:lagrangian}
\end{equation}
where, by~\eqref{eq:H},
\begin{equation}
2H-2c_0=-\Big(\psi'+\frac{\sin\psi}{\rho}+2c_0\Big),
\end{equation}
so that $\tfrac{k}{2}\rho(2H-2c_0)^{2}$ is the bending energy density per $2\pi$ and $\Sigma\rho$ the corresponding area term. The multipliers $\lambda_\rho$ and $\lambda_z$ are the momenta conjugate to $\rho$ and $z$.

Since $\endens$ depends on $z$ only through $z'$, the Euler--Lagrange equation $\delta\endens/\delta z=0$ gives $\lambda_z'=0$: $\lambda_z=f_z$ is constant, the resultant of the axial force transmitted across a parallel. For a particle held by adhesion alone no external force acts along the axis, hence $f_z=\lambda_z=0$: the vesicle translates freely to balance the forces, instead of being pinned to a distant substrate. A prescribed axial load, as in optical-trapping experiments~\cite{Necks2025}, would correspond to $f_z\neq0$.

The Euler--Lagrange equations for $\psi$ and $\rho$ are obtained in the standard way, by inserting the axisymmetric parametrization into the energy functional (see Eqs.~(3.2)--(3.5) of Ref.~\cite{SeifertBerndlLipowsky1991}, where $P=0$, $\lambda_\rho/k=\gamma$ and $C_0=-2c_0$ in our notation). For $f_z=0$ the equilibrium equations become
\begin{align}
\psi'' &= \frac{\cos\psi}{\rho}\Big(\frac{\sin\psi}{\rho}-\psi'\Big)
          +\frac{\lambda_\rho}{k\rho}\sin\psi ,
\label{eq:shape}\\[2pt]
\lambda_\rho' &= \frac{k}{2}\Big[(\psi'+2c_0)^{2}
          -\frac{\sin^{2}\!\psi}{\rho^{2}}\Big]
          +\Sigma ,
\label{eq:stress}
\end{align}
together with the geometric relations~\eqref{eq:frame}. In~\eqref{eq:stress}, $\Sigma$ is the multiplier of the area constraint of~\eqref{eq:augmented}. A planar membrane transmits the lateral tension $\Sigma+2kc_0^{2}$, and not $\Sigma$ (see~\eqref{eq:Seff}).

Because $\endens$ does not depend explicitly on the arclength $s$, the system admits a Beltrami (Hamiltonian) first integral, which is conserved along the meridian. It lowers the order of the shape equations and, at the free end $s_c$, enters the transversality condition that yields the adhesion law stated below.

The contact line $s=s_c$ is a free boundary~\cite{NapoliGoriely2020}. There the meridian joins the cap continuously,
\begin{equation}
\rho(s_c)=a\sin\thetac,\qquad \psi(s_c)=\thetac ,
\label{eq:contactBC}
\end{equation}
and, the endpoint being free, the variation with respect to $s_c$ furnishes in addition the natural transversality condition of the adhesion problem. It expresses a local balance between the adhesion energy and the jump of the meridional curvature across the contact line~\cite{Deserno2004,capovilla2002stresses},
\begin{equation}
\frac{k}{2}\big(\psi'(s_c)-1/a\big)^{2}=\Delta\gamma .
\label{eq:adhesion}
\end{equation}
Physically, the adhesion energy released at the contact line is paid by a discontinuity of the meridional curvature, of modulus $\sqrt{2\Delta\gamma/k}=\sqrt2\,\eta/a$. The condition does not involve the global multipliers; it prescribes the meridional curvature of the free membrane at the contact line and, together with~\eqref{eq:contactBC} and the boundary conditions at infinity, selects the wrapping angle $\thetac$: the energy of \S\ref{sec:energy}--\ref{sec:landscape} is evaluated at prescribed $\thetac$, and its stationarity with respect to $\thetac$ reproduces~\eqref{eq:adhesion} at the order retained.

\section{The wrapping neck as an elastic boundary layer}
\label{sec:asymptotics}
In the limit $\eps\to0$ the particle reduces to a point at the pole, and the membrane relaxes, over the macroscopic scale $\Rmem$, to its unperturbed outer shape. We regard the mean curvature $H_0$ of this shape at the pole and the spontaneous curvature $c_0$ as prescribed, with $aH_0$ and $ac_0$ of order $\eps$, so that both are comparable with the inverse macroscopic scale $1/\Rmem$. Near the contact line, however, the meridian must pass from the slope $\psi=\thetac$ of the bound cap to the gentle slope of the outer membrane within a meridional distance of order $a\ll\Rmem$; this region is an elastic boundary layer. We follow the matched-asymptotic analysis of inclusion-induced boundary layers of Ref.~\cite{BiscariNapoli2007} and the standard techniques~\cite{hinch1991perturbation}.

To resolve the layer we rescale the arclength and the radius to the particle scale,
\begin{equation}
\xi=\frac{s-s_c}{a},\qquad \varrho=\frac{\rho}{a},\qquad
\Lambda=\frac{a\,\lambda_\rho}{k},
\label{eq:stretch}
\end{equation}
By~\eqref{eq:eps} and the definitions of the dimensionless groups, the
ambient curvatures scale as $c_0a=\eps\,\sigma_0$ and $H_0a=\eps\,\Hb$, whereas
the tension scales as $\Sigma a^{2}/k=\Order{\eps^{2}}$. Consequently, in these variables the inner counterpart of~\eqref{eq:shape}--\eqref{eq:stress} reduces to the system
\begin{align}
\ddot\psi &= \frac{\cos\psi}{\varrho}\Big(\frac{\sin\psi}{\varrho}-\dot\psi\Big)
             +\frac{\Lambda}{\varrho}\sin\psi,
\label{eq:inner_shape}\\[2pt]
\dot\Lambda &= \frac12\Big[(\dot\psi+2\eps\sigma_0)^{2}
             -\frac{\sin^{2}\!\psi}{\varrho^{2}}\Big]
             +\Order{\eps^{2}},
\label{eq:inner_stress}
\end{align}
where dots denote $\dd/\dd\xi$, together with $\dot\varrho=\cos\psi$ and
with the inner mean curvature $2\meanc\equiv-(\dot\psi+\sin\psi/\varrho)$, so
that $H=(1/a)\,\meanc$.

The remainder $\Order{\eps^{2}}$ in~\eqref{eq:inner_stress} is the contribution of the tension. Equation~\eqref{eq:inner_shape} has no remainder: for $P=0$ and $f_z=0$ the axial multiplier $\lambda_z$ vanishes, so that the inner problem contains only $\Lambda=a\lambda_\rho/k$, the dimensionless radial membrane stress, and the neck transmits no axial force to the outer membrane.

Continuity with the bound cap imposes the contact values
\begin{equation}
\varrho(0)=\sin\thetac,\qquad \psi(0)=\thetac ,
\label{eq:contact}
\end{equation}

whereas the matching to the outer membrane (\S\ref{sub:matching7}) supplies the far-field condition. Two features of the problem govern the analysis that follows. First, the tension does not affect the neck at leading and at first order, so that, up to the order required, the boundary layer reduces to a problem of bending and geometry alone. Second, the inner expansion is regular, with $\Lambda=\Order{\eps}$: at leading order this follows from the matching (\S\ref{sec:catenoid}), and it is consistent at the next order (\S\ref{sec:jacobi}),
\begin{equation}
\psi=\Psi_0+\eps\,\Psi_1+\Order{\eps^{2}},\qquad
\varrho=\varrho_0+\eps\,\varrho_1+\Order{\eps^{2}},\qquad
\Lambda=\eps\,\Lambda_1+\Order{\eps^{2}}.
\label{eq:innerexp}
\end{equation}

\subsection{Leading order: the catenoidal neck}
\label{sec:catenoid}
At $\Order{1}$, \eqref{eq:inner_shape}--\eqref{eq:inner_stress} admit the first integral $\tfrac12\varrho_0(\dot\Psi_0^{2}-\sin^{2}\!\Psi_0/\varrho_0^{2})+\Lambda_0\cos\Psi_0=c_1$, that is $(\varrho_0\Lambda_0)^{\boldsymbol\cdot}=c_1$. Matching to the outer membrane, where $\Lambda_0\to0$, $\Psi_0\to0$ and $\varrho_0\dot\Psi_0\to0$ with $\varrho_0\sim\xi$ as $\xi\to\infty$, gives $c_1=0$, hence $\Lambda_0=c_2/\varrho_0$. Then $\varrho_0^{2}\dot\Lambda_0=-c_2\cos\Psi_0\to-c_2$, whereas $\varrho_0^{2}\dot\Lambda_0=\tfrac12[(\varrho_0\dot\Psi_0)^{2}-\sin^{2}\!\Psi_0]\to0$: thus $c_2=0$, $\Lambda_0\equiv0$ and $\dot\Psi_0=\pm\sin\Psi_0/\varrho_0$. The upper sign is a closed sphere, which cannot match the outer membrane; the lower sign gives
\begin{equation}
\dot\Psi_0=-\frac{\sin\Psi_0}{\varrho_0}
\qquad\Longleftrightarrow\qquad \meanc_0=0 .
\label{eq:minimal}
\end{equation}
The leading inner surface is therefore minimal. Using
$\dot\varrho_0=\cos\Psi_0$, one verifies the first integral
\begin{equation}
\varrho_0\sin\Psi_0=\varrho_n=\sin^{2}\!\thetac ,
\label{eq:firstintegral}
\end{equation}
where the constant is fixed by the contact data~\eqref{eq:contact}. Its geometric meaning is the following: the meridian is vertical, $\sin\Psi_0=1$, at the throat, where the radius attains its minimum, so that the constant is the throat radius $\varrho_n$. The leading neck is therefore a catenoid, with its throat at the axial position $z_n$, whose meridian reads
\begin{equation}
\varrho_0=\varrho_n\cosh\!\Big(\frac{z-z_n}{a\varrho_n}\Big) .
\label{eq:catenoid}
\end{equation}

As $\xi\to\infty$ the catenoid~\eqref{eq:catenoid} is unbounded, so that~\eqref{eq:firstintegral} gives $\sin\Psi_0=\varrho_n/\varrho_0\to0$ and therefore $\dot\varrho_0=\cos\Psi_0\to1$; consequently $\varrho_0\sim\xi$, $\Psi_0\simeq\sin\Psi_0$, and we obtain the far-field tails
\begin{equation}
\varrho_0(\xi)\sim\xi,\qquad
\Psi_0(\xi)\sim\frac{\varrho_n}{\xi}\qquad(\xi\to\infty),
\label{eq:cattail}
\end{equation}
which match to the outer field in \S\ref{sub:matching7}. Since $c_0$, $\Sigma$ and $P$ enter only at higher order, this leading surface coincides with the neck of the tension-clamped problem~\cite{DesernoBickel2003,Deserno2004}. Because $\meanc_0=0$ and $ac_0=\Order{\eps}$, the Helfrich density~\eqref{eq:helfrich} is $\Order{\eps^{2}}$ pointwise, so that $W_{\mathrm{layer}}^{(0)}=0$; the constant $2kc_0^{2}$ is absorbed into the effective tension (\S\ref{sec:reservoir}), while higher-order curvature effects are evaluated in \S\ref{sec:energy}.

\subsection{The first correction: the Jacobi problem of the catenoid}
\label{sec:jacobi}
Because the leading surface stores no energy, the bending energy of the layer is set by the first correction, which we write as a normal displacement $\eps w(\xi)$ of the catenoid (the logarithm comes instead from the tension acting on the excess area of the catenoid, \S\ref{sub:compenergy}). With $2\meanc=-(\dot\psi+\sin\psi/\varrho)$, a positive variation of the mean curvature lowers the slope, so that
\begin{equation}
\Psi_1=-\dot w .
\label{eq:PsiW}
\end{equation}
Linearizing~\eqref{eq:inner_shape}--\eqref{eq:inner_stress} about the catenoid, with $\Lambda=\eps\Lambda_1$ as in~\eqref{eq:innerexp}, gives the first-order mean curvature, $2\meanc=\eps\,\Jcat[w]$, where $\Jcat\equiv\Delta_{\mathrm{cat}}+|\II|^{2}$ is the Jacobi operator of the minimal surface~\cite{osserman1986survey}, that is, the linearization of the mean curvature under normal displacements. On the catenoid, with $\dd s^{2}=\dd\xi^{2}+\varrho_0^{2}\,\dd\varphi^{2}$, one has $\Delta_{\mathrm{cat}}f=\ddot f+(\cos\Psi_0/\varrho_0)\dot f$ on axisymmetric functions and, by~\eqref{eq:minimal} and~\eqref{eq:firstintegral}, $|\II|^{2}=2\sin^{2}\!\Psi_0/\varrho_0^{2}=2\varrho_n^{2}/\varrho_0^{4}=-2K$. Hence
\begin{equation}
\Jcat[w]=\ddot w+\frac{\cos\Psi_0}{\varrho_0}\,\dot w
+\frac{2\sin^{2}\!\Psi_0}{\varrho_0^{2}}\,w .
\label{eq:jacobiexplicit}
\end{equation}

\emph{Homogeneous solutions.} With $\varrho_0=\varrho_n\cosh u$, $u=(z-z_n)/(a\varrho_n)$ (\S\ref{sec:catenoid}), the metric is conformal, $\dd s^{2}=\varrho_n^{2}\cosh^{2}\!u\,(\dd u^{2}+\dd\varphi^{2})$, and
\begin{equation}
\Jcat=\frac{1}{\varrho_n^{2}\cosh^{2}\!u}
\Big(\frac{\dd^{2}}{\dd u^{2}}+\frac{2}{\cosh^{2}\!u}\Big) ,
\label{eq:jacobiconformal}
\end{equation}
so that $\Jcat f=0$ reads
\begin{equation}
\frac{\dd^{2}f}{\dd u^{2}}+\frac{2}{\cosh^{2}\!u}\,f=0 ,
\label{eq:homog}
\end{equation}
with the elementary solutions
\begin{equation}
f_1(u)=\tanh u ,\qquad f_2(u)=u\tanh u-1 .
\label{eq:jacobimodes}
\end{equation}
They are the normal displacements generated by a translation along the axis and by a change of the throat radius. Any combination of them is a neighbouring minimal surface to first order, with vanishing first-order mean curvature; consequently, the cost of the neck arises only from the particular solution forced by the source.

\emph{The source.} The linearized problem is of fourth order, since $\Jcat$ acts twice on $w$. On the layer $2\meanc=\eps\,\Jcat[w]$ and, in the inner variables, $2c_0=2\eps\sigma_0$; consequently, to $\Order{\eps^{2}}$ the Helfrich density $(2\meanc-2c_0)^{2}$ becomes $\eps^{2}(\Jcat[w]-2\sigma_0)^{2}$, and the layer energy is the quadratic form $\int\varrho_0\,(\Jcat[w]-2\sigma_0)^{2}\,\dd\xi$ in the first variation of $2\meanc$ (assembled in \S\ref{sec:energy}). Since $\Jcat$ is self-adjoint with respect to $\varrho_0\,\dd\xi$, stationarity of this form gives
\begin{equation}
\Jcat\big[\,\Jcat[w]-2\sigma_0\,\big]=0 ,
\label{eq:fourthorder}
\end{equation}
from which it follows that $\Jcat[w]-2\sigma_0=c_3f_1+c_4f_2$. The neck joins a membrane of mean curvature $H_0$ at the pole, that is, $\Jcat[w]\to2\Hb$ as $u\to\infty$. Since $f_2\sim u$ is unbounded, $c_4=0$, and since $f_1\to1$, $c_3=2(\Hb-\sigma_0)$; the forced Jacobi equation is therefore
\begin{equation}
\Jcat[w]=2\sigma_0+2(\Hb-\sigma_0)\tanh u .
\label{eq:forcing}
\end{equation}
The ambient curvature thus enters through the source and not through the multiplier: linearization of~\eqref{eq:inner_stress} gives
\begin{equation}
\dot\Lambda_1=+\frac{\sin\Psi_0}{\varrho_0}\,\big(\Jcat[w]-2\sigma_0\big) .
\label{eq:Lambda1}
\end{equation}
Since $\Jcat$ annihilates the homogeneous modes, $\Lambda_1$ is independent of their amplitudes $c_5,c_6$, so that its value at the contact line plays no role. These amplitudes are fixed by the contact conditions $w(u_c)=\dot w(u_c)=0$, that is $\varrho_1(0)=\Psi_1(0)=0$, with $\varrho_1=w\sin\Psi_0$. Apart from the constant $2\sigma_0$, the source depends on the data only through $\Hb-\sigma_0$, the departure of the ambient mean curvature from the preferred one. If this difference vanishes, the far field of the correction contains no logarithm, and for $\sigma_0=\Hb=0$ the neck is a catenoid to first order, $\Psi_1\equiv0$.

\emph{Far field.} The equation~\eqref{eq:forcing} is integrated by variation of parameters, the Wronskian being $f_1f_2'-f_1'f_2=1$. In the conformal parameter,
\begin{equation}
\frac{\dd^{2}w}{\dd u^{2}}+\frac{2}{\cosh^{2}\!u}\,w=\mathcal G ,
\qquad \mathcal G=\varrho_n^{2}\cosh^{2}\!u\,\big[2\sigma_0+2(\Hb-\sigma_0)\tanh u\big] .
\label{eq:forcedODE}
\end{equation}
Since $\varrho_0=\varrho_n\cosh u\sim\xi$,
\begin{equation}
u\simeq\ln\frac{2\xi}{\varrho_n}\qquad(\xi\to\infty),
\label{eq:uxi}
\end{equation}
whereas $f_1\to1$ and $f_2\sim\ln\xi$: the mode that changes the throat radius grows as the logarithm of the distance. By~\eqref{eq:forcing},
\begin{equation}
\mathcal G(u)=\varrho_n^{2}\big[\,\sigma_0+\sigma_0\cosh 2u
+(\Hb-\sigma_0)\sinh 2u\,\big]
=\varrho_n^{2}\Big[\,\sigma_0+\tfrac12\Hb\,e^{2u}\Big]+\Order{e^{-2u}} .
\label{eq:Gexp}
\end{equation}
The growing part is the ambient curvature: it produces $w\simeq\tfrac12\Hb\,\xi^{2}$, the quadratic growth of the height of a surface of mean curvature $H_0$ measured from its pole, and contributes to the potential term of~\eqref{eq:forcing} the additional constant $-\Hb\varrho_n^{2}$. The constant part of the source is then, in effect, $\varrho_n^{2}(\sigma_0-\Hb)$, and forces the secular solution $w\simeq\tfrac12\varrho_n^{2}(\sigma_0-\Hb)\,u^{2}\sim\ln^{2}\xi$. Consequently, by~\eqref{eq:PsiW} and $\dd u/\dd\xi=1/\varrho_0$, differentiation lowers by one the power of $\xi$ that multiplies each power of the logarithm, and
\begin{equation}
\Psi_1\sim-\Hb\,\varrho_0+\frac{c_7}{\varrho_0}+B\,\frac{\ln\varrho_0}{\varrho_0},
\qquad B=(\Hb-\sigma_0)\,\sin^{4}\!\thetac
\qquad(\xi\to\infty),
\label{eq:psi1tail}
\end{equation}
where $c_7$ is fixed by the inner contact conditions and the ambient slope $-\Hb\varrho_0$ corresponds to $2\meanc\to2\eps\Hb$. The expansion is written in $\varrho_0$ and not in $\xi$: the two differ by a constant, $\varrho_0=\xi+\sin\thetac\cos\thetac+o(1)$, which for $\Hb\neq0$ would add the term $-\Hb\sin\thetac\cos\thetac$ to the tail in $\xi$. For $\Hb=0$ the two forms coincide to the order considered. Both $B$ and the mean-curvature tail $\Jcat[w]-2\Hb\simeq-(\Hb-\sigma_0)\,\varrho_n^{2}/\varrho^{2}$ vanish with the mismatch $\Hb-\sigma_0$. Because $B$ enters the energy linearly through the cross term with $c_0$, its sign relative to $\Psi_0=\varrho_n/\xi>0$ matters; for $\Hb=0$ it gives $B=-\sigma_0\sin^{4}\!\thetac$, confirmed numerically in Fig.~\ref{fig:c0tests}b. By~\eqref{eq:uxi}, $u$ grows as $\ln\xi$, which is the origin of the logarithm in the energy of the layer (\S\ref{sec:energy}).

\emph{The case $\Hb=\sigma_0$.} For $\Hb=\sigma_0$, that is $H_0=c_0$ in physical variables, the right-hand side of~\eqref{eq:forcing} is the constant $2\sigma_0$. The correction is nevertheless nontrivial for $\sigma_0\neq0$. By~\eqref{eq:minimal} and~\eqref{eq:jacobiexplicit} one has $\Jcat[\tfrac12\varrho_0^{2}]=2$, so that
\[
w=\tfrac12\,\sigma_0\,\varrho_0^{2}+c_5f_1+c_6f_2 ,
\]
with $c_5,c_6$ fixed by the contact conditions, since the Wronskian of $f_1,f_2$ does not vanish. For $\sigma_0=\Hb=0$ the contact data are homogeneous, so that $c_5=c_6=0$, $w\equiv0$ and $\Psi_1\equiv0$: the neck is the catenoid to first order. The energy of this case is discussed at the end of \S\ref{sec:energy}.

\section{The composite solution and the energy of the neck}
\label{sec:energy}
We now match the inner solution of \S\ref{sec:asymptotics} to the outer field of a membrane coupled to a tension reservoir~\cite{DesernoBickel2003,Deserno2004,ACL2015,ACL2017}, construct the composite solution to first order in $c_0$, and evaluate the neck energy. Here the tension is a prescribed control parameter.

\subsection{Effective tension and outer scales}\label{sec:reservoir}
For a planar membrane ($\psi=0$), \eqref{eq:stress} yields $\lambda_\rho'=\Sigma+2kc_0^{2}$. The lateral tension transmitted by the flat far field is therefore the effective tension
\begin{equation}
\Seff=\Sigma+2kc_0^{2},\qquad \bar\Sigma_{\rm eff}=\frac{\Seff\Rmem^{2}}{k}=\frac{\Sigma\Rmem^{2}}{k}+2\sigma_0^{2},
\label{eq:Seff}
\end{equation}
rather than the multiplier $\Sigma$ of~\eqref{eq:augmented}. The constant density $2kc_0^{2}$ of the Helfrich energy is thus absorbed into $\Seff$, which acts on the excess area of the membrane (see~\eqref{eq:Eref} below).

\emph{Conventions and scaling.} In the reservoir ensemble at zero excess pressure ($P=0$), $\Seff$ (or $\bar\Sigma_{\rm eff}$) is prescribed at fixed $c_0$. It governs the outer screening length $\ell_{\mathrm{sc}}$ of~\eqref{eq:screening}, the far-field condition $\lambda_\rho\to\Seff\rho$ (Appendix~\ref{sub:numerics}), and the excess-area cost. We expand in $\eps$ at fixed $\sigma_0=ac_0/\eps$, $\bar\Sigma_{\rm eff}=\Seff a^{2}/(k\eps^{2})$, and $\eta=\Order{1}$ (so that the elastocapillary length $\ell_{\mathrm{ec}}=a/\eta=\sqrt{k/\Delta\gamma}$ is of order $a$), matching the ordering of Refs.~\cite{ACL2016,ACL2017}.

\subsection{The deflection charge and the outer field}\label{sub:outer}
\emph{Overlap geometry.} By~\eqref{eq:cattail}, $\Psi_0\simeq\varrho_n/\varrho$ for $\varrho\gg1$. In physical coordinates $\rho=a\varrho$, the meridian is a graph $z(\rho)$ with slope
\begin{equation}
\frac{\dd z}{\dd\rho}\simeq\tan\Psi_0\simeq\Psi_0
=\frac{a\sin^{2}\!\thetac}{\rho},
\label{eq:slope}
\end{equation}
giving the harmonic profile
\begin{equation}
z(\rho)\simeq a\sin^{2}\!\thetac\,\ln\frac{\rho}{a}+\mathrm{const} ,
\label{eq:zeta0}
\end{equation}
whose amplitude $a\sin^{2}\!\thetac$ we call the \emph{deflection charge}, the strength of the particle as a source of the outer field (for a bound particle as a source of stress, see Ref.~\cite{muller2005geometry}). Its excess area over the flat projection, $\dd A-\dd A_\perp\simeq\tfrac12|\nabla z|^{2}\,2\pi\rho\,\dd\rho$, diverges logarithmically across the overlap:
\begin{equation}
\delta A_{\mathrm{neck}}
=\int\tfrac12|\nabla z|^{2}\,\dd A_\perp
=\pi\!\int_a^{\Rmem}\!\Big(\frac{a\sin^{2}\!\thetac}{\rho}\Big)^{2}\rho\,\dd\rho
=\pi a^{2}\sin^{4}\!\thetac\,\ln\frac1\eps\,\big(1+o(1)\big) .
\label{eq:excessarea}
\end{equation}

Expanding~\eqref{eq:helfrich} and using~\eqref{eq:Seff}, the energy of the free membrane relative to the flat state with the same projected area $A_\perp$ reads
\begin{equation}
W_{\mathrm{free}}=\frac{k}{2}\!\int(2H)^{2}\dd A\;-\;2kc_0\!\int2H\,\dd A\;+\;\Seff\,(A-A_\perp) ,
\label{eq:Eref}
\end{equation}
so that the constant density $2kc_0^{2}$ is absorbed into $\Seff$ and acts on the excess area $A-A_\perp$.

In the outer region $\Omega$ the membrane is nearly flat ($H_0=0$, $\Hb=0$), so that $\psi\simeq z'$ and, by~\eqref{eq:H}, $2H\simeq-\nabla^{2}z$, with $\nabla^{2}$ the Laplacian of the plane. In this Monge gauge~\cite{Seifert1997,deserno2015fluid}, the energy~\eqref{eq:Eref} linearizes to
\begin{equation}
W_{\mathrm{out}}=\int_{\Omega}\Big[\tfrac{k}{2}(\nabla^{2}z)^{2}
+\tfrac{\Seff}{2}|\nabla z|^{2}\Big]\,\dd A .
\label{eq:outer_energy}
\end{equation}
Since $2kc_0\nabla^{2}z$ is a divergence, it only contributes a boundary flux (see \S\ref{sub:compenergy}) and drops out of the Euler--Lagrange equation
\begin{equation}
k\,\nabla^{4}z-\Seff\,\nabla^{2}z=0 ,
\label{eq:outer_pde}
\end{equation}
with the tension screening length, that is, the length $\ell_{\mathrm{sc}}$ beyond which the tension suppresses the deformation of the membrane,
\begin{equation}
\ell_{\mathrm{sc}}=\sqrt{k/\Seff}=\frac{\Rmem}{\sqrt{\bar\Sigma_{\rm eff}}},\qquad
\frac{\ell_{\mathrm{sc}}}{a}=\frac{1}{\eps\sqrt{\bar\Sigma_{\rm eff}}} .
\label{eq:screening}
\end{equation}
The general axisymmetric solution of $\nabla^{2}(\nabla^{2}-\ell_{\mathrm{sc}}^{-2})z=0$ is
\begin{equation}
z(\rho)=c_8+c_9\ln\rho+c_{10}\,I_0(\rho/\ell_{\mathrm{sc}})+c_{11}\,K_0(\rho/\ell_{\mathrm{sc}}) ,
\label{eq:outer_general}
\end{equation}
where $I_0,K_0$ are modified Bessel functions~\cite{abramowitz1972handbook}. Planar decay as $\rho\to\infty$ requires $c_{10}=0$, and finite tension energy in~\eqref{eq:outer_energy} requires $c_9=0$. Setting $c_8=z_\infty$ and $c_{11}=-q$, we obtain
\begin{equation}
z(\rho)=z_\infty-q\,K_0(\rho/\ell_{\mathrm{sc}}),\qquad
z'(\rho)=\frac{q}{\ell_{\mathrm{sc}}}K_1(\rho/\ell_{\mathrm{sc}})\xrightarrow[\rho\ll\ell_{\mathrm{sc}}]{}\frac{q}{\rho},
\label{eq:outer_sol}
\end{equation}
with $K_1=-K_0'$. Since $K_0(x)\sim-\ln(x/2)-\gamma_E$ as $x\to0$, the near-field behaviour is $z\simeq q\ln\rho+\mathrm{const}$, in agreement with~\eqref{eq:zeta0}.

\subsection{Matching and the composite solution}\label{sub:matching7}
The inner ($\rho\ll\ell_{\mathrm{sc}}$) and outer ($\rho\gg a$) expansions overlap in the intermediate range
\begin{equation}
a\ll\rho\ll\ell_{\mathrm{sc}} ,\qquad\text{or}\qquad
1\ll\xi\ll\frac{1}{\eps\sqrt{\bar\Sigma_{\rm eff}}} ,
\label{eq:overlap}
\end{equation}
which is non-empty provided $a\ll\ell_{\mathrm{sc}}$, i.e.\ $\bar\Sigma_{\rm eff}\ll\eps^{-2}$ (for $\eps\sqrt{\bar\Sigma_{\rm eff}}=\Order{1}$ the two scales merge). Matching the inner slope~\eqref{eq:slope} to the outer slope~\eqref{eq:outer_sol} in~\eqref{eq:overlap} fixes the deflection charge
\begin{equation}
q=a\,\sin^{2}\!\thetac ,
\label{eq:qmatch}
\end{equation}
Since $q=\Order{a}$, $(z-z_\infty)/\Rmem=\Order{\eps}$, consistent with~\eqref{eq:outer_energy}; the screening replaces $\Rmem$ by $\ell_{\mathrm{sc}}$ in~\eqref{eq:excessarea}, shifting $\ln(\ell_{\mathrm{sc}}/a)=\ln(1/\eps)-\tfrac12\ln\bar\Sigma_{\rm eff}$ by $\Order{1}$.

\emph{First order in $c_0$.} Linearizing~\eqref{eq:shape}--\eqref{eq:stress} about $\psi=0$ gives $\lambda_\rho'=\Seff+2kc_0\psi'+\Order{\psi^{2}}$, whence $\lambda_\rho=\Seff\rho+2kc_0\psi+\Order{\psi^{2}}$ and
\begin{equation}
k\Big(\psi''+\frac{\psi'}{\rho}-\frac{\psi}{\rho^{2}}\Big)-\Seff\,\psi=\frac{2kc_0}{\rho}\,\psi^{2},
\label{eq:outer1}
\end{equation}
cubic terms in $\psi$ belonging to the inner problem. Expanding $\psi=\psi_0+c_0\psi_1$ with $\psi_0=(q/\ell_{\mathrm{sc}})K_1(x)$ and $x=\rho/\ell_{\mathrm{sc}}$, the decaying solution of~\eqref{eq:outer1} is
\begin{equation}
\psi_1=\frac{q^{2}}{\ell_{\mathrm{sc}}}\Big[\,Y(x)+\Big(\alpha-\ln\frac{\ell_{\mathrm{sc}}}{a}\Big)K_1(x)\Big],
\label{eq:psi1out}
\end{equation}
where $Y$ is the decaying solution of $Y''+Y'/x-(1+x^{-2})Y=2K_1^{2}(x)/x$ normalized by $xY(x)+\ln x\to0$ as $x\to0$. For $\rho\ll\ell_{\mathrm{sc}}$,
\begin{equation}
\rho\,\psi=q+c_0q^{2}\Big[\,\alpha-\ln\frac{\rho}{a}\Big]+\dots
\label{eq:rhopsi}
\end{equation}
Matching~\eqref{eq:rhopsi} to the inner tail~\eqref{eq:psi1tail} at $\Hb=0$ fixes $\alpha$ from the contact conditions $w(u_c)=\dot w(u_c)=0$. Variation of parameters in~\eqref{eq:forcedODE} yields $\dd(w/(\sigma_0\varrho_n^{2}))/\dd u\to u-u_c-\tfrac12\tan^{2}(\thetac/2)$ as $u\to\infty$, with $e^{u_c}=\cot(\thetac/2)$, so that
\begin{equation}
\alpha=\ln\frac{\sin^{2}\!\thetac}{2}+\ln\cot\frac{\thetac}{2}+\frac12\tan^{2}\frac{\thetac}{2},
\label{eq:alpha}
\end{equation}
which agrees with the numerical integration of~\eqref{eq:forcedODE} for $30^\circ\le\thetac\le150^\circ$ (the composite is tested in Appendix~\ref{sub:numerics}, item~(iv)); for $\Hb=0$, $\eps w/\cos\Psi_0$ also reproduces numerically the inner solution $z_{i2}$ of Ref.~\cite{BiscariNapoli2007}, Eq.~(2.7). Thus the outer field sees the effective charge $q+c_0q^{2}(\alpha-\ln(\ell_{\mathrm{sc}}/a))$, and the composite solution, obtained by the standard additive rule of matched asymptotics (inner plus outer solution, minus their common part in the overlap~\eqref{eq:overlap})~\cite{hinch1991perturbation}, reads
\begin{equation}
\psi_{\rm c}=\psi_{\rm in}+\psi_{\rm out}-\frac{q+c_0q^{2}\,[\alpha-\ln(\rho/a)]}{\rho},
\label{eq:composite}
\end{equation}
with $\psi_{\rm in}=\Psi_0+\eps\Psi_1$ and $\psi_{\rm out}=\psi_0+c_0\psi_1$.

\subsection{The energy of the composite}\label{sub:compenergy}
\emph{Zeroth order in $c_0$.} At $c_0=0$, the catenoid has zero bending energy, and the outer bending energy $\tfrac{k}{2}\!\int(2H)^{2}\dd A=\tfrac{\pi}{2}\,\Seff q^{2}$ (with $2H=qK_0(\rho/\ell_{\mathrm{sc}})/\ell_{\mathrm{sc}}^{2}$) is finite. The logarithmic cost, which is the value of the functional $W_{\mathrm{out}}$ of~\eqref{eq:outer_energy} on the outer field, comes entirely from $\Seff$ acting on the excess area~\eqref{eq:excessarea} up to $\rho\sim\ell_{\mathrm{sc}}$:
\begin{equation}
W_{\mathrm{out}}\simeq\Seff\,\delta A_{\mathrm{neck}}
=\pi\Seff\,q^{2}\ln\!\frac{\ell_{\mathrm{sc}}}{a}
=\pi k\,\bar\Sigma_{\rm eff}\,\eps^{2}\sin^{4}\!\thetac\,\ln\!\frac1\eps
\;\big(1+\Order{1/\ln}\big),
\label{eq:tensionenergy}
\end{equation}
the finite $\thetac$-dependent part being derived in \S\ref{sub:landscape7}. For small penetration, $\tau=1-\cos\thetac$ (the variable $z$ of Ref.~\cite{Deserno2004}), one has $\sin^{4}\!\thetac\simeq4\tau^{2}$ and, with $\tilde\Sigma=\Seff a^{2}/k$, $\ln(\ell_{\mathrm{sc}}/a)=-\tfrac12\ln\tilde\Sigma$, so that $W_{\mathrm{out}}/\pi k\simeq-2\tilde\Sigma\tau^{2}\ln\tilde\Sigma$: this is the logarithmic term of Eq.~(21) of Ref.~\cite{Deserno2004}, whose other terms are the finite part (\S\ref{sub:tense}). At finite angle, Eq.~(20) of the same reference, obtained in the small-gradient approximation, gives instead the coefficient $\sin^{4}\!\thetac/\cos^{2}\thetac$, because it imposes the slope $\tan\thetac$ at the contact line and thus the deflection charge $a\sin^{2}\!\thetac/\cos\thetac$, whereas the catenoid gives $q=a\sin^{2}\!\thetac$. The two coincide as $\thetac\to0$; at $\thetac=60^\circ$ they differ by a factor $4$, and the coefficient $\sin^{4}\!\thetac$ is the one confirmed by the full shape equations (Appendix~\ref{sub:numerics}).

\emph{First order in $c_0$.} Since $W_{\mathrm{free}}$ in~\eqref{eq:Eref} is stationary with respect to $\psi$ at fixed contact data and $\Seff$, the envelope theorem gives $\partial_{c_0}W=-2k\!\int_{\mathcal S_{\mathrm f}}2H\,\dd A$, whence
\begin{equation}
W(c_0)-W(0)=-2k\int_0^{c_0}\!\dd c'\int_{\mathcal S_{\mathrm f}(c')}\!2H\,\dd A ,
\label{eq:Ec0}
\end{equation}
where $\mathcal S_{\mathrm f}(c')$ is the equilibrium free surface at spontaneous curvature $c'$, with the same contact data and $\Seff$. Using $2H=-(\psi'+\sin\psi/\rho)$, $\rho'=\cos\psi$, and $\rho\psi\to0$ as $s\to s_{\max}$, we have the exact identity
\begin{equation}
\int_{\mathcal S_{\mathrm f}}2H\,\dd A=2\pi\rho_c\theta_c-2\pi\!\int_{s_c}^{s_{\max}}(\sin\psi-\psi\cos\psi)\,\dd s ,
\label{eq:meanidentity}
\end{equation}
with $\rho_c=a\sin\thetac$. On the catenoid ($H=0$), the integral in~\eqref{eq:meanidentity} equals $\rho_c\theta_c-q$, recovering the outer flux $2\pi q=2\pi a\sin^{2}\!\thetac$ (both $\rho_c\theta_c$ and $q$ vanish as $\thetac\to\pi$). Differentiating~\eqref{eq:meanidentity} with respect to $c_0$ at fixed contact data gives $-2\pi\!\int\psi\sin\psi\,\partial_{c_0}\psi\,\dd s$. For $\rho\gg a$, $\psi\sin\psi\simeq q^{2}/\rho^{2}$ and $\partial_{c_0}\psi\simeq q^{2}(\alpha-\ln(\rho/a))/\rho$ by~\eqref{eq:rhopsi}, so the integrand decays as $\ln\rho/\rho^{3}$. The integral converges without depending on $\ell_{\mathrm{sc}}$, yielding
\begin{equation}
W(c_0)=W(0)-4\pi k\,a\,c_0\sin^{2}\!\thetac+k\,c_0^{2}\,W_2(\thetac)+\dots ,
\label{eq:Eexpansion}
\end{equation}
where $W_2=\Order{a^{2}}$ is independent of $\ln(\ell_{\mathrm{sc}}/a)$. At leading order in $\tilde\Sigma$ the second variation of~\eqref{eq:Eref} at $c_0=0$ is the bending energy of the first-order mean curvature, since $2H=0$ on the catenoid, and stationarity of the quadratic form gives $kc_0^{2}W_2=-\tfrac{k}{2}\int(2H_1)^{2}\dd A$, minus the first-order bending energy of the layer. The first term of~\eqref{eq:densities} integrated from $u_c$ to $\infty$, with $e^{-u_c}=\tan(\thetac/2)$, gives $W_2=-2\pi a^{2}\sin^{4}\!\thetac\tan^{2}(\thetac/2)=-2\pi a^{2}\tau^{3}(2-\tau)$, in agreement with the shape equations (Appendix~\ref{sub:numerics}). At fixed $\Seff$ the spontaneous curvature produces a flux term of order $\eps$ and nothing at order $\eps^{2}\ln(1/\eps)$; the coefficient of the logarithm is that of $c_0=0$, with $\Seff$ in place of the tension.

\emph{Result.} Truncating at $\ell_{\mathrm{sc}}=\sqrt{k/\Seff}$, where $\ln(\ell_{\mathrm{sc}}/a)=\ln(1/\eps)+\Order{1}$, the logarithmic neck energy is
\begin{equation}
W_{\mathrm{neck}}
=\pi\,\Seff\,q^{2}\ln\frac{\ell_{\mathrm{sc}}}{a}
=\pi k\,\Gamma\,\sin^{4}\!\thetac\;\eps^{2}\ln\!\frac{1}{\eps}
 \;+\;\Order{\eps^{2}}\,k ,
\qquad
\Gamma=\bar\Sigma_{\rm eff}\ge0 ,
\label{eq:mainresult}
\end{equation}
up to relative corrections of order $\eps\ln(1/\eps)$ due to the scale dependence of the charge. The remainder $\Order{\eps^{3}\ln^{2}(1/\eps)}$ of~\eqref{eq:landscape_exact} is estimated from the mixed term in $c_0$ and $\Seff$: the flux $\int2H\,\dd A$ has a correction of relative order $\tilde\Sigma\ln^{2}\tilde\Sigma$, which multiplies the term in $c_0$ below. This is an estimate at the leading logarithm, not a complete derivation; it is consistent with the deviation of the flux from its limit, below $0.4\%$ at $\Seff=10^{-3}$ (Fig.~\ref{fig:c0tests}a). To it one adds the flux $-4\pi kac_0\sin^{2}\!\thetac$ of~\eqref{eq:Eexpansion}, which carries no logarithm. The coefficient is non-negative as a consequence of the sign of the tension, $\Seff>0$ (that is, $\Sigma>-2kc_0^{2}$, assumed throughout), and not of a property of the bending energy of the neck. In terms of the multiplier, $\Gamma=\Sigma\Rmem^{2}/k+2\sigma_0^{2}$: the term $2\sigma_0^{2}$ is the $2kc_0^{2}$ density of the preferred-curvature term acting as lateral tension on the excess area, and not a bending mechanism of the neck. The result is not a local property of the layer: the logarithms of the layer, listed below, cancel in the outer field, and the residual coefficient depends on the closure. For a closed vesicle with $H_0\neq c_0$ it has to be recomputed from the outer field (\S\ref{sec:summary}); the case $H_0=c_0$ is treated at the end of this section.

The factor $\sin^{4}\!\thetac$ differs from the $1-\cos\thetac$ of the cap, and both it and its second derivative vanish at $\thetac=0$: the neck does not move the onset of wrapping, but modifies the energy at intermediate angles, where the transition between shallow and deep wrapping occurs.

\subsection{Check by integration of the energy density}\label{sub:direct}
The result can be checked, and the structure of the layer made explicit, by integrating the energy density over the layer and the overlap.

\emph{Setting.} In the stretched variables the energy of the layer is exactly
\begin{equation}
W_{\mathrm{layer}}
=\frac{k}{2}\int_{\mathrm{neck}}(2H-2c_0)^{2}\,\dd A
=\pi k\int_{0}^{\xi_\ast}\varrho\,\big(2\meanc-2\eps\sigma_0\big)^{2}\,\dd\xi ,
\label{eq:innerenergy}
\end{equation}
with $\xi_\ast$ a cut in the overlap, $1\ll\xi_\ast\ll1/\eps$. By~\eqref{eq:innerexp} and $2\meanc_0=0$ one has $2\meanc=\eps\,\Jcat[w]+\Order{\eps^{2}}$, with $\Jcat[w]$ given by~\eqref{eq:jacobiexplicit}; we have checked this against the exact mean curvature of the normally displaced catenoid. Therefore, to leading order $\Order{\eps^{2}}$ at fixed $\xi_\ast$,
\begin{equation}
W_{\mathrm{layer}}
=\pi k\,\eps^{2}\int_{0}^{\xi_\ast}
   \varrho_0\,\big(\Jcat[w]-2\sigma_0\big)^{2}\,\dd\xi ,
\label{eq:quadenergy}
\end{equation}

\emph{The layer integral.} By~\eqref{eq:forcing}, $\Jcat[w]-2\sigma_0=2(\Hb-\sigma_0)\tanh u$. Since $\varrho_0=\varrho_n\cosh u$ and $\dd\xi=\varrho_n\cosh u\,\dd u$, the integrand of~\eqref{eq:quadenergy} reduces to $4(\Hb-\sigma_0)^{2}\varrho_n^{2}\sinh^{2}u\,\dd u$, and
\begin{equation}
\int\varrho_0\,(\Jcat[w]-2\sigma_0)^{2}\,\dd\xi
=(\Hb-\sigma_0)^{2}\varrho_n^{2}\,\big[\sinh2u-2u\big]_{u_c}^{u_\ast},
\label{eq:innerint}
\end{equation}
where $u_c$ labels the contact line, $\cosh u_c=1/\sin\thetac$, and $u_\ast$ the cut $\varrho_0(u_\ast)=\varrho_\ast$ in the overlap. With $\sinh2u_\ast=2\varrho_\ast^{2}/\varrho_n^{2}-1+\Order{\varrho_\ast^{-2}}$ and $u_\ast=\ln(2\varrho_\ast/\varrho_n)+\Order{\varrho_\ast^{-2}}$,
\begin{equation}
\int\varrho_0\,(\Jcat[w]-2\sigma_0)^{2}\,\dd\xi
=2(\Hb-\sigma_0)^{2}\varrho_\ast^{2}\;-\;2(\Hb-\sigma_0)^{2}\varrho_n^{2}\ln\varrho_\ast\;+\;(\Hb-\sigma_0)^{2}c_{12}(\thetac)+o(1),
\label{eq:innercut}
\end{equation}
with $c_{12}$ independent of $\varrho_\ast$. The first term is the Helfrich energy of the ambient membrane on a flat disc of radius $\varrho_\ast$, which any reference-state subtraction removes. The coefficient of the logarithm is \emph{negative}, and is the sum of two pieces: the ambient energy density acting on the excess area~\eqref{eq:excessarea}, $+2(\Hb-\sigma_0)^{2}\varrho_n^{2}\ln\varrho_\ast$, and the cross term between the ambient mismatch $2(\Hb-\sigma_0)$ and the decaying part of the curvature, $\Jcat[w]-2\Hb\simeq-4(\Hb-\sigma_0)e^{-2u}$, which gives twice as much with the opposite sign, $-4(\Hb-\sigma_0)^{2}\varrho_n^{2}\ln\varrho_\ast$.

For the reservoir, $\Hb=0$, so that $\Hb-\sigma_0=-\sigma_0$, and $\Jcat[w]=2\sigma_0(1-\tanh u)$, and the three terms of~\eqref{eq:Eref} have the densities per unit $u$, in units of $\pi k\eps^{2}$,
\begin{equation}
\text{bending: } 4\sigma_0^{2}\varrho_n^{2}e^{-2u},\qquad
\text{cross term: } -4\sigma_0^{2}\varrho_n^{2}\,(1+e^{-2u}),\qquad
\text{tension: } \bar\Sigma_{\rm eff}\,\varrho_n^{2}\,(1+e^{-2u}).
\label{eq:densities}
\end{equation}
The bending term is integrable; the other two grow as $u\simeq\ln\varrho$, and if the layer were the whole system the logarithm would have the coefficient $(\bar\Sigma_{\rm eff}-4\sigma_0^{2})\varrho_n^{2}$.

The cross term contains no $\ln(1/\eps)$ in total, by~\eqref{eq:meanidentity}, independently of the structure of the layer. The negative logarithm $-4\sigma_0^{2}\varrho_n^{2}\ln\varrho$ of~\eqref{eq:densities} must therefore be cancelled, and the cancellation is not local. In the overlap $\psi=(q-c_0q^{2}\ln(\rho/a))/\rho$ by~\eqref{eq:psi1tail}, in the orientation in which $q>0$, and $2H=c_0q^{2}/\rho^{2}$. The flux of the slope through the circle of radius $\rho$, $2\pi\rho\psi=2\pi(q-c_0q^{2}\ln(\rho/a))$, varies logarithmically with $\rho$, by $-c_0q^{2}\ln(\ell_{\mathrm{sc}}/a)$ up to the scale $\ell_{\mathrm{sc}}$ of the outer field. The outer solution is matched to the charge $q-c_0q^{2}\ln(\ell_{\mathrm{sc}}/a)$, not to $q$, as in~\eqref{eq:psi1out}, and contributes at $\rho\sim\ell_{\mathrm{sc}}$ a logarithm of opposite sign to the one accumulated by the cross term in the overlap. The numerical solution exhibits this compensation (Fig.~\ref{fig:c0tests}c).

\subsection{The case of equal curvatures}\label{sub:equalcurv}
For $\Hb=\sigma_0$ the coefficient $B$ of~\eqref{eq:psi1tail} vanishes and so does the layer integral~\eqref{eq:innerint}, which is proportional to $(\Hb-\sigma_0)^{2}$. With the solution $w$ of \S\ref{sec:jacobi} the first-order surface has $2\meanc=2\eps\sigma_0$, that is $2H=2c_0$: its mean curvature equals the spontaneous one, and the Helfrich density is of higher order than $\eps^{2}$. Without the correction ($w=0$) the layer would carry the density $2kc_0^{2}$ over the area of the catenoid, which is not zero and contains the logarithm of the excess area~\eqref{eq:excessarea}; for $\Hb=\sigma_0$ the correction removes it entirely. For $\sigma_0=\Hb=0$ the neck is the catenoid to first order, and the only energy of the neck at order $\eps^{2}\ln(1/\eps)$ is that of the tension on its excess area, the result~\eqref{eq:mainresult} with $\sigma_0=0$, that is, the flat membrane under tension treated in Ref.~\cite{Deserno2004}. For $\Hb=\sigma_0\neq0$ the outer surface must have $H_0=c_0$ and the plane is excluded: the natural closure is the sphere of radius $1/|c_0|$ (with $c_0<0$ in the convention of \S\ref{sec:model}, where $2H=-2/R$ on a sphere of radius $R$), whose Helfrich density vanishes and which is an equilibrium for the pressure $p=2\Sigma|c_0|$. For a normal perturbation $\upsilon$ its second variation is
\[
\tfrac12\!\int\!\Big[k\,(\Delta_{\mathrm{sph}}\upsilon+2c_0^{2}\upsilon)^{2}+\Sigma\big(|\nabla\upsilon|^{2}-2c_0^{2}\upsilon^{2}\big)\Big]\dd A ,
\]
without cross term. Away from the particle the outer field is the Green function of $-k\Delta_{\mathrm{sph}}+\Sigma-2kc_0^{2}$ on the sphere, with $\upsilon\simeq-q\ln s$ near the particle and $q=a\sin^{2}\!\thetac$; then $\Delta_{\mathrm{sph}}\upsilon+2c_0^{2}\upsilon=(\Sigma/k)\upsilon$, the bending term is finite, and the only logarithm is that of the tension term. For $\Rmem=1/|c_0|$, so that $\eps=a|c_0|$ and $\Gamma=\Sigma\Rmem^{2}/k=\Sigma/(kc_0^{2})=\Order{1}$,
\[
W_{\mathrm{neck}}=\pi k\,\Gamma\,\sin^{4}\!\thetac\;\eps^{2}\ln\frac1\eps+\Order{\eps^{2}}k ,
\]
that is $\Gamma=\bar\Sigma_{\rm eff}-2\sigma_0^{2}$ with $\sigma_0^{2}=1$: the part $2\sigma_0^{2}$ of $\bar\Sigma_{\rm eff}$, the density $2kc_0^{2}$ on the excess area, is removed by the correction $w$, and no flux of order $\eps$ appears. For $c_0\to0$ the expression has the form of the flat case. The coefficient of the logarithm was checked on the linearised sphere and, independently, on the full nonlinear shape equations of a closed vesicle (unit sphere, particle at a pole, fixed $\Sigma$ and pressure $p=2\Sigma|c_0|$, $30^\circ\le\thetac\le150^\circ$, $\Gamma=0.3$ and $1$, $0.002\le\eps\le0.05$): the grand potential relative to the bare sphere, after removal of the bending energy $4\pi k(1-\eps)^{2}(1-\cos\thetac)$ of the bound cap, contains no term of order $\eps^{0}$ or $\eps$ and its coefficient of $\eps^{2}\ln(1/\eps)$ equals $\pi k\Gamma\sin^{4}\!\thetac$ within $0.3\%$, whereas $\bar\Sigma_{\rm eff}$ in its place would give a coefficient larger by the factor $1+2/\Gamma$.

\section{Wrapping energy, thresholds and hysteresis}\label{sec:landscape}
\subsection{The reservoir wrapping energy}\label{sub:landscape7}
Expanding~\eqref{eq:helfrich} gives $\tfrac{k}{2}(2H-2c_0)^{2}=\tfrac{k}{2}(2H)^{2}-2kc_0(2H)+2kc_0^{2}$, where the constant $2kc_0^{2}$ is absorbed into $\Seff$ by~\eqref{eq:Seff}. The curvature-coupling term $-2kc_0\!\int2H\,\dd A$ is linear in the deformation, and we evaluate it separately on the cap and on the free membrane. On the bound cap ($2H=-2/a$, area $2\pi a^{2}(1-\cos\thetac)$), it yields $\int_{\mathcal S_{\mathrm b}}\!2H\,\dd A=-4\pi a(1-\cos\thetac)$. On the free membrane, in the Monge gauge $2H\simeq-\nabla^{2}z$, the integral reduces to the boundary flux $-2\pi[\,\rho\,z'\,]$. Since $\rho\,z'\to q$ at the inner edge and $\rho\,z'\to0$ as $\rho\to\infty$ by~\eqref{eq:outer_sol}, we obtain
\begin{equation}
\int_{\mathcal S_{\mathrm f}}\!2H\,\dd A=2\pi q=2\pi a\sin^{2}\!\thetac ,
\label{eq:totalH}
\end{equation}
so that the total mean curvature is
\begin{equation}
\int_{\mathcal S}\!2H\,\dd A=-4\pi a(1-\cos\thetac)+2\pi a\sin^{2}\!\thetac
=-2\pi a\,(1-\cos\thetac)^{2}.
\label{eq:totalHsum}
\end{equation}
This is $\Order{\thetac^{4}}$ for shallow wrapping, rather than $\Order{\thetac^{2}}$ as on the cap alone: the contributions of the cap and of the neck cancel at order $\thetac^{2}$, and $c_0$ enters the energy through $(1-\cos\thetac)^{2}$, as the tension does in~\eqref{eq:capexcess} (verified numerically in Appendix~\ref{sub:numerics}).

\emph{Neck term of order $\eps$.} For a mother membrane of ambient mean curvature $H_0$, the linear variation of the Helfrich density is $k(2H_0-2c_0)\,\delta(2H)$. Integrating $\delta(2H)$ over $\mathcal S_{\mathrm f}$ via~\eqref{eq:totalH} gives
\begin{equation}
W^{(1)}_{\mathrm{neck}}=4\pi k\,a\,(H_0-c_0)\sin^{2}\!\thetac=4\pi k\,a\,(H_0-c_0)\,\tau\,(2-\tau),
\label{eq:neckflux}
\end{equation}
which reproduces the unbound-membrane term of Ref.~\cite{ACL2017}, Eqs.~(27) and (31)--(32) (with $M\to H_0$, $m\to c_0$). In the reservoir ($H_0=0$), combining $-4\pi k\,a\,c_0\sin^{2}\!\thetac$ with the cap term $8\pi k\,a\,c_0(1-\cos\thetac)$ yields the $\Order{\eps}$ contribution $4\eps\sigma_0\tau^{2}$. Thus, although the catenoid has zero Helfrich energy, the neck stores energy at $\Order{\eps}$ through the linear coupling~\eqref{eq:neckflux}, whereas the quadratic deformation cost enters only at $\Order{\eps^{2}\ln(1/\eps)}$ via~\eqref{eq:mainresult}.

Adding the bound-cap bending, the adhesion energy and the tension energy $W_{\Seff}$ to the $\Order{\eps}$ terms above, the wrapping energy relative to the flat state reads (in units of $\pi k$, with $\tau=1-\cos\thetac\in[0,2]$)
\[
\frac{W_{\mathrm{wrapp}}}{\pi k}=4\tau\Big[1-\tfrac12\eta^{2}\Big]+4\eps\sigma_0\tau^{2}+\frac{W_{\Seff}}{\pi k}.
\]

\emph{Tension energy.} The membrane pays $\Seff$ per unit excess area $A-A_\perp$ drawn from the reservoir. On the bound cap ($\rho=a\sin\psi$, $\dd s=a\,\dd\psi$), the excess area over the projected disc is
\begin{equation}
\delta A_{\mathrm b}
=2\pi\!\int_{0}^{\thetac}\!\rho\,(1-\cos\psi)\,\dd s
=2\pi a^{2}\!\int_{0}^{\thetac}\!\sin\psi\,(1-\cos\psi)\,\dd\psi
=\pi a^{2}(1-\cos\thetac)^{2} ,
\label{eq:capexcess}
\end{equation}
which is $\Order{\thetac^{4}}$ as $\thetac\to0$. The finite excess area of the neck is of the same order and becomes essential for $\thetac>\pi/2$, where the neck overhangs the cap. On the catenoid, with $\tanh u_c=\cos\thetac$ ($u_c<0$ for $\thetac>\pi/2$), the area from $u_c$ to the projected radius $\rho_\ast=a\varrho_n\cosh u_\ast$ is $\pi a^{2}\varrho_n^{2}\,[u+\tfrac12\sinh2u]_{u_c}^{u_\ast}$. Adding the cap area $2\pi a^{2}\tau$ and subtracting $\pi\rho_\ast^{2}$ gives, for all $\thetac$,
\begin{equation}
\delta A(\rho_\ast)=2\pi a^{2}(1-\cos\thetac)-\pi a^{2}\sin^{2}\!\thetac\cos\thetac
+\pi a^{2}\sin^{4}\!\thetac\Big[\ln\frac{2\rho_\ast}{a}+\ln\frac{1-\cos\thetac}{\sin^{3}\thetac}-\frac12\Big],
\label{eq:exactarea}
\end{equation}
up to terms vanishing as $\rho_\ast/a\to\infty$. Near complete wrapping ($\chi=\pi-\thetac\ll1$), the first two terms give $4\pi a^{2}+\Order{\chi^{4}}$, with no $\Order{\chi^{2}}$ term.

In the reservoir the cut-off is removed by the outer screening. Beyond the overlap radius $\rho_\ast$, the outer solution~\eqref{eq:outer_sol} carries the bending-plus-tension energy $\pi\Seff q^{2}\,xK_0(x)K_1(x)$ ($x=\rho_\ast/\ell_{\mathrm{sc}}$), since $\tfrac{\dd}{\dd x}[xK_0K_1]=-x(K_0^{2}+K_1^{2})$. Expanding for $x\ll1$ gives $\pi\Seff q^{2}[\ln(2\ell_{\mathrm{sc}}/\rho_\ast)-\gamma_E]$. Adding the inner free-membrane excess area $\delta A(\rho_\ast)-\delta A_{\mathrm b}$ and the cap excess~\eqref{eq:capexcess}, we find that the cut-off $\rho_\ast$ cancels, and we obtain
\begin{equation}
W_{\Seff}=\pi\Seff a^{2}\,\mathcal A(\thetac),\qquad
\mathcal A=\tau^{2}(3-\tau)+\sin^{4}\!\thetac\Big[\ln\frac{4\ell_{\mathrm{sc}}}{a}-\gamma_E-\frac12+\ln\frac{1-\cos\thetac}{\sin^{3}\thetac}\Big],
\label{eq:tensionexact}
\end{equation}
where $\gamma_E=0.5772\ldots$ is the Euler constant and $\ln(4\ell_{\mathrm{sc}}/a)=\ln(1/\eps)+\ln4-\tfrac12\ln\bar\Sigma_{\rm eff}$. In units of $\pi k$, $W_{\Seff}=\bar\Sigma_{\rm eff}\eps^{2}\mathcal A(\thetac)$, and the full landscape is
\begin{equation}
\frac{W_{\mathrm{wrapp}}}{\pi k}=4\tau\Big[1-\tfrac12\eta^{2}\Big]+4\eps\sigma_0\tau^{2}-2\eps^{2}\sigma_0^{2}\,\tau^{3}(2-\tau)+\bar\Sigma_{\rm eff}\eps^{2}\mathcal A(\thetac)+\Order{\eps^{3}\ln^{2}(1/\eps)}.
\label{eq:landscape_exact}
\end{equation}
Equation~\eqref{eq:tensionexact} agrees with the full nonlinear shape equations for $20^\circ\le\thetac\le170^\circ$ within $0.16\%$ at $\Seff a^{2}/k=10^{-3}$ and $0.03\%$ at $10^{-4}$ (Appendix~\ref{sub:numerics}). We retain the non-logarithmic $\Order{\eps^{2}}$ terms in~\eqref{eq:landscape_exact} because they are the only tension effects surviving at $\thetac=\pi$, where the neck logarithm~\eqref{eq:tensionenergy} vanishes.

\subsection{Partial-wrapping threshold}\label{sub:partial}
As $\thetac\to0$, $\tau=\thetac^{2}/2+\Order{\thetac^{4}}$, while $\tau^{2}(3-\tau)=\Order{\thetac^{4}}$ and $\sin^{4}\!\thetac\,\ln[\tau/\sin^{3}\!\thetac]=\Order{\thetac^{4}\ln\thetac}$. Thus both $4\eps\sigma_0\tau^{2}$ and $\bar\Sigma_{\rm eff}\eps^{2}\mathcal A$ vanish at $\thetac=0$ along with their first two derivatives, leaving the second variation
\begin{equation}
\frac{1}{\pi k}\,\frac{\dd^{2}W_{\mathrm{wrapp}}}{\dd\thetac^{2}}\bigg|_{\thetac=0}=4-2\eta^{2}.
\label{eq:secondvar}
\end{equation}
The unwrapped state $\thetac=0$ is therefore locally stable for $\eta^{2}<2$ and loses stability at the classical binding threshold~\cite{LipowskyDoebereiner1998,DesernoBickel2003,Deserno2004}, at which, for $c_0=0$, the adhesion energy per unit area equals the bending energy density $2k/a^{2}$ of the bound cap,
\begin{equation}
\eta_{W}^{2}=2 .
\label{eq:etaW}
\end{equation}
Its independence of tension~\cite{DesernoBickel2003,Deserno2004} and spontaneous curvature~\cite{ACL2015,ACL2017} follows directly from~\eqref{eq:capexcess} and from the cap--neck curvature cancellation~\eqref{eq:totalHsum}, which is derived here from the shape equations through~\eqref{eq:meanidentity}.

\subsection{Complete wrapping: stability and equal energy}\label{sub:wrappedstab}
Near complete wrapping ($\chi=\pi-\thetac\ll1$), $\tau=2-\chi^{2}/2+\Order{\chi^{4}}$, $\tau^{2}(3-\tau)=4-\tfrac34\chi^{4}+\Order{\chi^{6}}$, $\tau^{3}(2-\tau)=4\chi^{2}+\Order{\chi^{4}}$, and the neck term is $\Order{\chi^{4}\ln\chi}$; consequently, the expansion of~\eqref{eq:landscape_exact} reads
\begin{equation}
\frac{W_{\mathrm{wrapp}}}{\pi k}=\frac{W_{\mathrm{wrapp}}(\pi)}{\pi k}+\big[\eta^{2}-2(1+2\eps\sigma_0)^{2}\big]\chi^{2}+\Order{\eps^{2}\chi^{4}\ln\chi},
\label{eq:wrappedstab}
\end{equation}
The coefficient of $\chi^{2}$ is the curvature of the landscape at closure, and its sign decides whether the wrapped state is stable. The connected, fully wrapped state is therefore locally stable for $\eta^{2}>\eta_{s}^{2}=2(1+2\eps\sigma_0)^{2}=2+8\eps\sigma_0+8\eps^{2}\sigma_0^{2}$; the tension does not shift this limit, because the area drawn from the reservoir is stationary at $\thetac=\pi$. This recovers the stability limit of Refs.~\cite{ACL2015,ACL2016,ACL2017}, where the narrow-neck expansion~\cite{ACL2016}, Eq.~(35), gives, to first order in $\eps\sigma_0$, $\delta E\simeq\pi k\sin^{2}\!\thetac\,(\eta^{2}-2-8\eps\sigma_0)$ for a flat membrane; the term in $\eps^{2}\sigma_0^{2}$ of~\eqref{eq:wrappedstab} completes the limit $2(1+2ac_0)^{2}$ of \S\ref{sec:summary}. The optimal neck radius is $q$.

At $\thetac=\pi$ ($\tau=2$, $\mathcal A=4$), the neck vanishes without topology change (neck fission, which would add the Gaussian contribution $4\pi\bar k$, is not considered), and~\eqref{eq:landscape_exact} reduces to
\begin{equation}
\frac{W_{\mathrm{wrapp}}(\pi)}{\pi k}=8-4\eta^{2}+16\eps\sigma_0+4\bar\Sigma_{\rm eff}\eps^{2}.
\label{eq:Wfw}
\end{equation}
Thus $W_{\mathrm{wrapp}}(\pi)<0$ above the equal-energy threshold
\begin{equation}
\eta_{\mathrm{fw}}^{2}=2+4\eps\sigma_0+\bar\Sigma_{\rm eff}\eps^{2}=\eta_W^{2}+4\eps\sigma_0+\bar\Sigma_{\rm eff}\eps^{2},
\label{eq:etafw}
\end{equation}
shifted by $4\eps\sigma_0$~\cite{ACL2015,ACL2017} and by the area $4\pi a^{2}$ drawn against $\Seff$. The equal-energy line differs from the stability limit $\eta_s^{2}=2(1+2\eps\sigma_0)^{2}$ of~\eqref{eq:wrappedstab}: in the tensionless limit with $\sigma_0>0$ and $2+4\eps\sigma_0<\eta^{2}<2(1+2\eps\sigma_0)^{2}$, the fully wrapped state has lower energy than the free state but is unstable against partial unwrapping, and a partially wrapped state is the global minimum.

\subsection{Comparison with the wrapping at low tension}\label{sub:tense}
For $c_0=0$ and $\tilde\Sigma=\Seff a^{2}/k=\bar\Sigma_{\rm eff}\eps^{2}$, \eqref{eq:landscape_exact} reduces to $W_{\mathrm{wrapp}}/\pi k=4\tau(1-\tfrac12\eta^{2})+\tilde\Sigma\mathcal A(\thetac)$, which we compare with Deserno~\cite{Deserno2004}.

\emph{(i) Free-membrane energy.} Expanding~\eqref{eq:tensionexact} for small $\tau$ gives $\tilde\Sigma[\mathcal A-\tau^{2}]=-2\tilde\Sigma\tau^{2}\,[\ln(\tilde\Sigma\tau/2)+2\gamma_E]+\Order{\tau^{3}\ln\tau}$, recovering Eq.~(21) of Ref.~\cite{Deserno2004}. Whereas Eq.~(21) is restricted to $\tau\ll1$ and at finite angle Ref.~\cite{Deserno2004} has numerical solutions and the small-gradient energy, Eq.~(20) (\S\ref{sub:compenergy}), \eqref{eq:tensionexact} is analytic for all $\thetac$, to leading order in $\tilde\Sigma$, (Fig.~\ref{fig:c0tests}e) and yields analytically the vanishing of the free-membrane energy at complete wrapping, $\mathcal A(\pi)-\tau^{2}(\pi)=0$.

\emph{(ii) Zero-energy line.} For $c_0=0$, Eq.~\eqref{eq:etafw} gives $\eta^{2}=2+\tilde\Sigma$, which coincides with the zero-energy line of Ref.~\cite{Deserno2004}, Eq.~(25), whose adhesion parameter is $2\eta^{2}$.

\emph{(iii) Envelopment line E.} Table~\ref{tab:deserno} compares the equal-energy line of partially and fully wrapped states from~\eqref{eq:landscape_exact}, from the full shape equations, and from Eq.~(26) of Ref.~\cite{Deserno2004}: all three agree within $8\times10^{-4}$ ($0.04\%$) up to $\tilde\Sigma=0.2$, so that for the line E the formula of Ref.~\cite{Deserno2004} is as accurate as~\eqref{eq:landscape_exact}. The error of~\eqref{eq:tensionexact} against the shape equations rises from $0.9\%$ at $\tilde\Sigma=10^{-2}$ to $10\%$ at $\tilde\Sigma=0.2$; the high-tension regime $\tilde\Sigma\gg1$ ($\ell_{\mathrm{sc}}\lesssim a$) lies outside the asymptotic overlap~\eqref{eq:overlap}. As a further check, for $\Seff=0.02\,$dyn/cm, $a=30\,$nm and $k=20\,k_BT$, that is $\tilde\Sigma\simeq0.22$, Ref.~\cite{DesernoBickel2003} reports a barrier of about $22\,k_BT$ at the envelopment line, and~\eqref{eq:landscape_exact} gives the same value.
{
\begin{table}[t]
\centering
\caption{Position of the envelopment line E in $\eta^{2}$ ($c_0=0$): equal energy of the partially and of the fully wrapped state, from Eq.~\eqref{eq:landscape_exact}, from the full nonlinear shape equations (interpolated, $\thetac\ge10^{\circ}$), and from Eq.~(26) of Ref.~\cite{Deserno2004}. Last column: zero-energy line of the fully wrapped state, $2+\tilde\Sigma$.}
\label{tab:deserno}
\begin{tabular}{lcccc}
\hline
$\tilde\Sigma$ & Eq.~\eqref{eq:landscape_exact} & shape equations & Ref.~\cite{Deserno2004}, Eq.~(26) & $2+\tilde\Sigma$\\
\hline
0.01 & 2.0102 & 2.0102 & 2.0102 & 2.0100\\
0.05 & 2.0513 & 2.0513 & 2.0512 & 2.0500\\
0.20 & 2.2085 & 2.2077 & 2.2078 & 2.2000\\
\hline
\end{tabular}
\end{table}}

\subsection{Ordering, bistability and hysteresis}\label{sub:hysteresis}
The landscape~\eqref{eq:landscape_exact} depends on $\varsigma_0=\eps\sigma_0=ac_0$ and $\tilde\Sigma=\bar\Sigma_{\rm eff}\eps^{2}=\Seff a^{2}/k$. For $\sigma_0=\Order{1}$, the $\Order{\eps}$ term $4\varsigma_0\tau^{2}$ dominates over the $\Order{\eps^{2}\ln(1/\eps)}$ tension term and governs the nature of the transition.

\emph{Spinodals.} Interior stationary points of~\eqref{eq:landscape_exact}, $0<\thetac<\pi$, exist only for $\eta^{2}-2=\Order{\eps,\tilde\Sigma}$, since otherwise the term $4\tau(1-\tfrac12\eta^{2})$ dominates; they satisfy $\partial_{\thetac}W_{\mathrm{wrapp}}=0$, that is
\begin{equation}
\eta^{2}(\thetac)=2+4\varsigma_0\tau-2\varsigma_0^{2}\tau^{2}(3-2\tau)+\frac{\tilde\Sigma}{2}\,\frac{\mathcal A'(\thetac)}{\sin\thetac},
\label{eq:spinodal}
\end{equation}
where $\mathcal A'=\dd\mathcal A/\dd\thetac$. For $\varsigma_0=0$ the limits of metastability are the points at which $\dd\eta^{2}/\dd\thetac=0$: the upper spinodal S1 is the maximum of~\eqref{eq:spinodal} for $\thetac<\pi$, where the partially wrapped minimum merges with the barrier and disappears, and the lower spinodal S2 is its minimum near $\thetac=\pi$, where the open-neck minimum disappears. They have no closed form and follow from~\eqref{eq:spinodal}.

\emph{The $\Order{\eps}$ landscape.} As $\tilde\Sigma\to0$, $W_{\mathrm{wrapp}}/\pi k=4\tau(1-\tfrac12\eta^{2})+4\varsigma_0\tau^{2}$~\cite{ACL2015,ACL2017}. For $\varsigma_0>0$, a single minimum $\tau=(\eta^{2}-2)/4\varsigma_0$ grows continuously from $0$ ($\eta^{2}=2$) to $2$ ($\eta^{2}=2+8\varsigma_0$). For $\varsigma_0<0$ the quadratic term is concave, and consequently the free ($\eta^{2}<2$) and the fully wrapped ($\eta^{2}>2+8\varsigma_0$) states coexist across a bistable window of width $8\eps|\sigma_0|$, separated by a barrier at $\tau=(\eta^{2}-2)/4\varsigma_0$ with equal energy at $\eta^{2}=2+4\varsigma_0$. Hysteresis driven solely by the logarithm occurs only near the multicritical point $\sigma_0=0$.

\begin{figure}[t]
\centering
\includegraphics[width=0.90\textwidth]{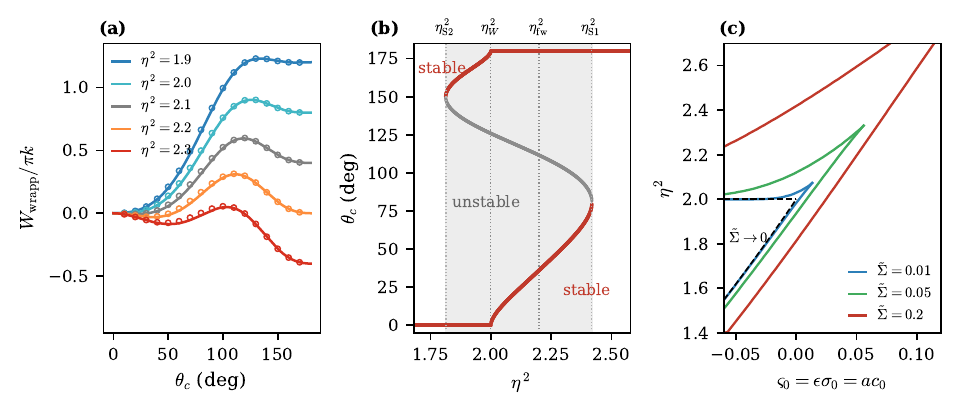}
\caption{Wrapping of a small particle by a membrane in a tension reservoir, from the landscape~\eqref{eq:landscape_exact}. (a)~$W_{\mathrm{wrapp}}/\pi k$ against the wrapping angle for $\tilde\Sigma=0.2$, $\varsigma_0=0$ and $\eta^{2}=1.9,\,2.0,\,2.1,\,2.2,\,2.3$: lines, Eq.~\eqref{eq:landscape_exact}; circles, full nonlinear shape equations. (b)~Minima (red, stable) and maxima (grey, unstable) of the landscape against $\eta^{2}$ ($\tilde\Sigma=0.2$, $\varsigma_0=0$); the shaded window, $1.81<\eta^{2}<2.42$, is that in which two minima coexist, bounded by the limits of metastability of the two states (the spinodals S1 and S2 of Ref.~\cite{Deserno2004}); the dotted lines mark $\eta^{2}_{\rm S2}$, $\eta_W^{2}=2$, $\eta^{2}_{\rm fw}$ (equal energy of the free and fully wrapped states) and $\eta^{2}_{\rm S1}$. The state at $180^{\circ}$ is stable for $\eta^{2}>2(1+2\varsigma_0)^{2}$, and below it the branch near $180^{\circ}$ is a state with an open neck. (c)~Lower and upper limits of the window in which two minima coexist against $\varsigma_0=ac_0$ for $\tilde\Sigma=0.01$, $0.05$ and $0.2$; the limits meet at $\varsigma_{\rm cr}$; dashed, $\tilde\Sigma\to0$: $2(1+2\varsigma_0)^{2}$ and $2$ for $\varsigma_0<0$.}
\label{fig:reservoir}
\end{figure}

\emph{Effect of the logarithm.} Figure~\ref{fig:reservoir}c shows the coexistence limits $\eta_{\rm S2}^{2}$ and $\eta_{\rm S1}^{2}$ against $\varsigma_0$. As $\tilde\Sigma\to0$ for $\varsigma_0<0$, they approach $2(1+2\varsigma_0)^{2}$ and $2$. For $\varsigma_0>0$, bistability survives only up to a critical value $\varsigma_{\rm cr}(\tilde\Sigma)=0.014,\,0.057,\,0.17$ for $\tilde\Sigma=0.01,\,0.05,\,0.2$, that is $\varsigma_{\rm cr}\simeq0.9$--$1.4\,\tilde\Sigma$. Consequently, the logarithm controls the hysteresis only when $|\varsigma_0|\lesssim\tilde\Sigma$, that is $|\sigma_0|\lesssim\Gamma\eps$.

\emph{Low-tension membrane at $c_0=0$.} For $\varsigma_0=0$ and $\tilde\Sigma=0.2$ ($\eps=0.1$, $\bar\Sigma_{\rm eff}=20$, Fig.~\ref{fig:reservoir}a,b), wrapping proceeds continuously above $\eta_W^{2}=2$ until the partially wrapped minimum disappears at the upper spinodal $\eta_{\rm S1}^{2}=2.42$ (equal energy, the line E of Table~\ref{tab:deserno}, is at $\eta_{\rm E}^{2}=2.21$). Decreasing $\eta^{2}$ from $\thetac=\pi$, the closed state becomes unstable below $\eta_s^{2}=2$ (the limit~\eqref{eq:wrappedstab} at $\varsigma_0=0$), where it bifurcates into a metastable open-neck minimum (e.g.\ $\thetac\simeq166^{\circ}$, neck radius $0.06\,a$ at $\eta^{2}=1.9$) that persists down to the lower spinodal $\eta_{\rm S2}^{2}=1.81$ ($1.81<\eta^{2}<2.41$ from the full shape equations). The thresholds $\eta_{\rm S2}^{2}<\eta_W^{2}<\eta_{\mathrm{fw}}^{2}<\eta_{\rm E}^{2}<\eta_{\rm S1}^{2}$ and the open-neck branch reproduce the spinodals S1, S2~\cite{DesernoBickel2003,Deserno2004} and the metastable state found numerically in Ref.~\cite{Deserno2004}, extending them to $c_0\neq0$ (Fig.~\ref{fig:reservoir}c).

\section{Concluding remarks and outlook}\label{sec:summary}
By treating the wrapping neck as an elastic boundary layer, we have derived the energetics of partial engulfment directly from the shape equations, without postulating a minimal surface: the catenoid and its first correction are the first two terms of the inner expansion. For the tension reservoir the expansion recovers results known in the literature: the binding threshold $\eta_W^{2}=2$, independent of the tension and of $c_0$~\cite{LipowskyDoebereiner1998,DesernoBickel2003,Deserno2004,ACL2015,ACL2017}; the stability limit $2(1+2\eps\sigma_0)^{2}$ of the wrapped state, distinct from the equal-energy line~\eqref{eq:etafw}~\cite{ACL2015,ACL2016,ACL2017}; the bistability window of width $8\eps|\sigma_0|$ for $\sigma_0<0$~\cite{ACL2015,ACL2017} (\S\ref{sub:hysteresis}); and the leading catenoid~\cite{Deserno2004} and the $\Order{\eps}$ slope-flux term~\cite{ACL2017}. In addition, it determines the next-order cost of the neck, of order $\eps^{2}\ln(1/\eps)$, which originates from the effective lateral tension acting on the logarithmic excess area of the catenoid. The resulting landscape~\eqref{eq:landscape_exact} describes the crossover between the low-tension regime of Ref.~\cite{Deserno2004} and the local landscape of Refs.~\cite{ACL2015,ACL2017}.

\emph{New results.} (i)~The composite solution for the tension reservoir. It joins the first correction to the catenoid (the inner problem of Ref.~\cite{BiscariNapoli2007}, here with free contact angle) to the outer field to first order in $c_0$, and fixes the constant~\eqref{eq:alpha} without cut-off (\S\ref{sec:energy}). (ii)~The energy of the neck beyond order $\eps$. The logarithmic term $\pi k\,\Gamma\sin^{4}\!\thetac\,\eps^{2}\ln(1/\eps)$, with $\Gamma=\bar\Sigma_{\rm eff}$ the effective tension acting on the excess area of the catenoid, and the finite part~\eqref{eq:tensionexact} are obtained analytically at every wrapping angle. For $c_0=0$ and small penetration they reduce to the result of Ref.~\cite{Deserno2004}, whose exact finite-angle energies are numerical; the angular dependence $\sin^{4}\!\thetac$, instead of the $1-\cos\thetac$ of the cap, governs the landscape between shallow and deep wrapping. (iii)~For $c_0\neq0$, the logarithm of the layer due to the curvature coupling, of coefficient $-4\sigma_0^{2}$ in~\eqref{eq:densities}, cancels in the outer field, so that the coefficient of the logarithm is not a local property of the layer and depends on the closure (\S\ref{sub:direct}, \S\ref{sub:equalcurv}). (iv)~The landscape~\eqref{eq:landscape_exact}, which contains as limits the description of Ref.~\cite{Deserno2004} and the local landscape of Refs.~\cite{ACL2015,ACL2017} and describes the crossover between them, including the spinodals~\cite{DesernoBickel2003} and the open-neck branch that Ref.~\cite{Deserno2004} finds numerically at $c_0=0$, here following from the closed-form landscape, also for $c_0\neq0$. For the reservoir the expansion also gives the order of the remainder in~\eqref{eq:landscape_exact}, estimated at the leading logarithm (\S\ref{sub:direct}): within the regular inner expansion the catenoidal neck generates no energy term of lower order than those retained. The neck energy is confirmed against the full nonlinear shape equations (Appendix~\ref{sub:numerics}).

\emph{Physical relevance.} The landscape depends on three numbers: the adhesion $\eta^{2}$, the curvature $\varsigma_0=ac_0$ and the dimensionless tension $\tilde\Sigma=\Seff a^{2}/k$. For $k\simeq20\,k_BT$ and $a=50\,$nm one has $\tilde\Sigma\simeq0.03$, $0.3$ and $3$ for $\Seff=10^{-6}$, $10^{-5}$ and $10^{-4}\,$N/m. The expansion rests on the overlap $a\ll\rho\ll\ell_{\mathrm{sc}}$, so that it requires $\ell_{\mathrm{sc}}\gtrsim a$; we take $\tilde\Sigma\lesssim0.2$ as its limit, where the error of~\eqref{eq:tensionexact} against the shape equations is $10\%$ (\S\ref{sub:tense}). This corresponds, for $a=50\,$nm, to $\Seff\lesssim6.6\times10^{-6}\,$N/m and, for a bead of radius $0.5\,\mu$m, to $\Seff\lesssim6.6\times10^{-8}\,$N/m. The theory therefore describes small particles on weakly tensed membranes; larger tensions lie outside the overlap on which the expansion rests, and for $\tilde\Sigma\gg1$ they belong to the high-tension regime of Ref.~\cite{Deserno2004}.

\emph{Hysteresis and its observability.} In units of $\pi k\simeq63\,k_BT$ the barrier on the equal-energy line E between the partially and the fully wrapped state at $\varsigma_0=0$ is, from~\eqref{eq:landscape_exact}, $0.028$, $0.11$ and $0.32$ for $\tilde\Sigma=0.01$, $0.05$ and $0.2$, that is about $2$, $7$ and $20\,k_BT$: the tension alone yields a barrier of several $k_BT$ only for $\tilde\Sigma\gtrsim0.05$, at the upper end of the range of validity. For $\varsigma_0<0$ the barrier at the equal-energy point is $4|\varsigma_0|\pi k\simeq250\,|\varsigma_0|\,k_BT$ for $\tilde\Sigma\to0$, with a window of width $8|\varsigma_0|$ in $\eta^{2}$: for $\varsigma_0=-0.05$, that is $c_0=-1\,\mu\mathrm{m}^{-1}$ at $a=50\,$nm, about $13\,k_BT$ and a width $0.4$. The bistability is therefore controlled by the spontaneous curvature of the bilayer, tunable by the lipid composition, rather than by the logarithm. The open-neck branch below $\eta_W^{2}$ (\S\ref{sub:hysteresis}) has a throat radius $0.06\,a\simeq3\,$nm at $a=50\,$nm, comparable to the bilayer thickness and at the limit of the continuum description: it is a property of the landscape, not a quantitative prediction.

\emph{Vesicles and bilayer asymmetry.} In a closed vesicle the mother membrane has a mean curvature $H_0\neq0$ at the contact, in the convention of \S\ref{sec:model}. With the cap term of~\eqref{eq:Ebend_split} and the neck flux~\eqref{eq:neckflux}, the part of the landscape of order $\eps$ is, in units of $\pi k$, $4\tau(1-\tfrac12\eta^{2}+2aH_0)+4a(c_0-H_0)\tau^{2}$, which reduces to~\eqref{eq:landscape_exact} for $H_0=0$; the subtraction of the energy $2k(H_0-c_0)^{2}A_{\mathrm b}$ of the segment of mother membrane that the cap replaces is of order $\eps^{2}$. The onset of adhesion is then $\eta_W^{2}=2+4aH_0$, shifted by the curvature of the mother membrane and not by $c_0$; the stability limit of the wrapped state is $2+8ac_0-4aH_0$ to first order about $\eta^{2}=2$, the expansion of the limit $\eta^{2}=2(1+2ac_0-aH_0)^{2}$ of Ref.~\cite{ACL2017}, and $c_0$ acts through $c_0-H_0$, whose sign decides whether the engulfment is continuous or bistable. These are the results of the local landscape of Refs.~\cite{ACL2015,ACL2017}: the effect of the bilayer asymmetry in a real vesicle is fixed by the local curvature before contact, which depends on the shape and on the tension of the vesicle.

\emph{Endocytosis.} Cellular endocytosis is an active process: actin dynamics counteract the membrane tension, which opposes the completion of the clathrin-coated pit~\cite{Boulant2011,MechanoCME}, and the neck is closed by dedicated proteins, so that the present model does not describe it. It describes the passive, elastic part of the uptake of small particles, and is the natural reference for experiments on vesicles~\cite{Dietrich1997,Spanke2020,OptTweez2023,Ayala2023} and nanoparticles~\cite{Strobl2014}. We do not compare with these data: the theory is axisymmetric, and the thermal fluctuations that determine the uptake energy in the optical-trap experiments of Ref.~\cite{Ayala2023} are not included. For its bead of radius $0.5\,\mu$m and $k=1.6$--$20\,k_{\mathrm B}T$, $\tilde\Sigma\lesssim0.2$ requires the stiffest membranes at tensions below about $6\times10^{-8}\,$N/m; its flaccid vesicles have $\tilde\Sigma>1$. The vesicles of Ref.~\cite{Spanke2020} are deflated, at fixed area and volume, with a tension compatible with zero (about $3\times10^{-10}\,$N/m at the most likely value); this is the vesicle closure of the Outlook rather than the reservoir.

\emph{Outlook.} Once the layer has been integrated out, the wrapped particle can be replaced by a point object at the pole carrying the self-energy $W_{\mathrm{neck}}(\thetac)$ of~\eqref{eq:mainresult} and~\eqref{eq:neckflux}. The two-scale free-boundary problem then reduces to the Helfrich energy $W_{\mathrm m}$ of the mother membrane, that is, of the membrane without the inclusion, plus one localized term. Let $\tau=1-\cos\thetac$, let $\eta^{2}$ be the adhesion parameter of \S\ref{sec:landscape}, and let $\Hb=aH_0/\eps$ and $\sigma_0=ac_0/\eps$ be the dimensionless mean curvatures of the mother membrane at the contact and of the spontaneous curvature, respectively. With $\Gamma$ the coefficient of the logarithm in~\eqref{eq:mainresult}, the effective energy is
\begin{equation*}
W_{\mathrm{eff}}=W_{\mathrm m}+\pi k\Big\{4\tau\big(1-\tfrac12\eta^{2}\big)+8\eps\Hb\,\tau+4\eps\,(\sigma_0-\Hb)\,\tau^{2}+\Gamma\,\eps^{2}\sin^{4}\!\thetac\,\ln\frac1\eps+\Order{\eps^{2}}\Big\} ,
\end{equation*}
where $W_{\mathrm m}$ is the energy of the mother membrane, the first two terms in braces are those of the cap and of the adhesion, the next two are the curvature terms of the cap and of the neck, of order $\eps$, and the last is the neck energy of order $\eps^{2}\ln(1/\eps)$.
The $\Order{1}$ and $\Order{\eps}$ terms recover the local landscape of Ref.~\cite{ACL2017}, Eqs.~(31)--(32), valid for a mother membrane of arbitrary shape and stated for $|\eta^{2}-2|=\Order{\eps}$, the range of the stability limits; our $\Order{1}$ term $4\tau(1-\tfrac12\eta^{2})$ holds for any $\eta^{2}$. The boundary layer adds the neck term of order $\eps^{2}\ln(1/\eps)$, a correction of relative order $\eps\ln(1/\eps)$ whose coefficient $\Gamma$ depends on the global closure.

The coefficient $\Gamma$ equals $\bar\Sigma_{\rm eff}$ in the reservoir, where the terms of order $\eps^{2}$ are given in full by~\eqref{eq:landscape_exact}, and equals $\Sigma\Rmem^{2}/k=\bar\Sigma_{\rm eff}-2\sigma_0^{2}$ on the sphere $H_0=c_0$ of \S\ref{sub:equalcurv}. For a general closed vesicle with $H_0\neq c_0$, where the far field carries a nonzero $\Hb$, $\Gamma$ must be computed from the outer field of that closure.

This formulation should make the engulfment in closed vesicles at fixed area and volume, as well as the curvature-mediated interactions between several bound particles or inclusions~\cite{Reynwar2007,MidyaAuthGompper2023}, analytically tractable. The same reduction can be expected to apply to the narrow necks of budding, whose limit shapes have an infinitesimal neck~\cite{JulicherLipowsky1993}. Finally, since at order $\eps$ the energy depends on the local mean curvature alone,~\eqref{eq:neckflux}, a partially wrapped particle is driven by the gradient of the mean curvature~\cite{ACL2017}, whereas at complete wrapping ($\tau=2$) the terms of order $\eps$ reduce to $16\eps\sigma_0$ and no longer depend on $\Hb$, in agreement with Ref.~\cite{ACL2017}. The migration of partially wrapped Janus particles toward sites of high deviatoric curvature on tense vesicles~\cite{Li2017Langmuir}, which is a different mechanism driven by the excess area, and any coupling of anisotropic inclusions to the curvature deviator~\cite{KraljIglic1999}, which is not small on the neck, where the principal curvatures are opposite, and which can stabilize narrow necks, are not included in the present model. With the values estimated in Ref.~\cite{KraljIglic1999}, inclusions aligned on a neck of radius of order $a$ lower the energy by about $k_{\mathrm B}T$, small against $4\pi k$ but comparable to $W_{\mathrm{neck}}$: an order-of-magnitude estimate, relevant only for strongly anisotropic inclusions.

\section*{Acknowledgments}
This work was carried out under the auspices of GNFM--INdAM. The author thanks the referees for their careful reports. The first referee drew attention to the literature on the wrapping of small particles that is cited in \S\ref{sec:intro}, and indicated the finite part of the excess area of the catenoid, Eq.~\eqref{eq:exactarea}, and the structure of the cross term analysed in \S\ref{sub:direct}.

\textbf{Declaration of generative AI use.} During the preparation of this version of the work, the author used Claude (Anthropic) to assist with drafting and editing the text, reviewing the literature, and formatting the figures. Following the use of this tool, the author carefully reviewed and edited the content as necessary and takes full responsibility for the content of the published article.

\appendix
\section{Numerical verification}\label{sub:numerics}
\emph{Setup.} To test the parameter-free predictions~\eqref{eq:mainresult} (with $\Gamma=\bar\Sigma_{\rm eff}$) and~\eqref{eq:qmatch}, we solve the full nonlinear shape equations~\eqref{eq:shape}--\eqref{eq:stress} by collocation~\cite{kierzenka2001bvp} in units $a=k=1$, with $P=0$ and control parameter $\Seff=\Sigma+2kc_0^{2}$ (so that $\ell_{\mathrm{sc}}/a=1/\sqrt{\Seff}$). Boundary conditions are $\rho=\sin\thetac$, $\psi=\thetac$ at $s=s_c$ and $\psi\to0$, $\lambda_\rho\to\Seff\rho$ as $s\to s_{\max}$, initialized from the screened catenoid. From the converged meridian we evaluate the free-membrane energy~\eqref{eq:Eref} relative to the flat state,
\begin{equation}
W_{\mathrm{full}}(\thetac,\Seff)
=\pi k\!\int_{s_c}^{s_{\max}}\!(2H)^{2}\rho\,\dd s
-4\pi kc_0\!\int_{s_c}^{s_{\max}}\!2H\,\rho\,\dd s
+2\pi\Seff\!\int_{s_c}^{s_{\max}}\!\rho\,(1-\cos\psi)\,\dd s ,
\label{eq:Efull}
\end{equation}
which excludes the closed-form $\Order{1}$ cap and adhesion terms of~\eqref{eq:landscape_exact} and isolates the neck contribution. The deflection charge is extracted from $\rho\,\psi\to q$ in the overlap.

\begin{figure}[t]
\centering
\includegraphics[width=0.52\textwidth]{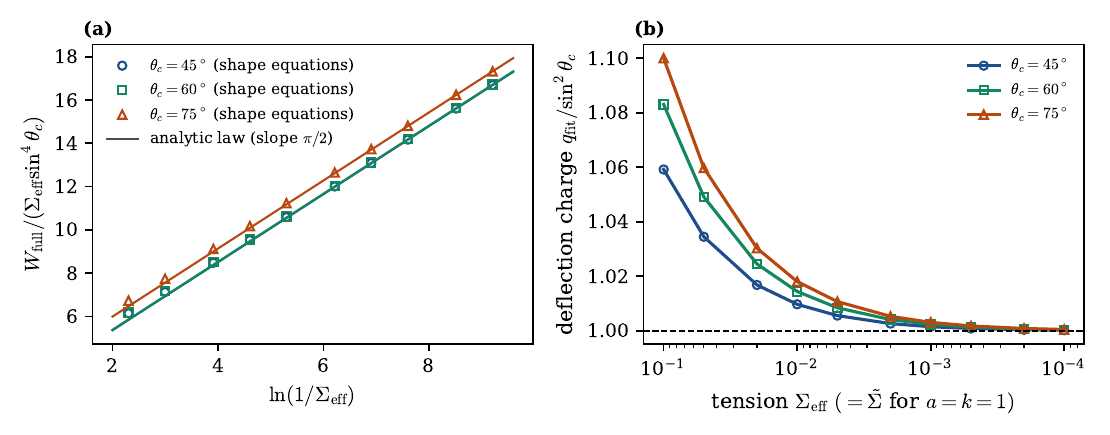}
\caption{Full nonlinear shape equations versus the law~\eqref{eq:verifexact} ($a=1$, $k=1$, $\sigma_0=0$; $\thetac=45^\circ,60^\circ,75^\circ$, $\Seff\in[10^{-4},10^{-1}]$). (a)~Compensated energy of the free membrane $W_{\mathrm{full}}/(\Seff\sin^{4}\!\thetac)$ against $\ln(1/\Seff)$: symbols, shape equations; lines, the law~\eqref{eq:verifexact}, of slope $\pi/2$, the coefficient of the logarithm of~\eqref{eq:mainresult}, with an intercept that depends on $\thetac$ (the lines for $45^\circ$ and $60^\circ$ nearly coincide). (b)~Deflection charge extracted from the far field, $q_{\mathrm{fit}}/\sin^{2}\!\thetac\to1$ as $\Seff\to0$, confirming the matched value $q=a\sin^{2}\!\thetac$ of~\eqref{eq:qmatch}.}
\label{fig:verification}
\end{figure}

\emph{Predictions and results at $c_0=0$} (Fig.~\ref{fig:verification}). For $c_0=0$, $\Gamma=\bar\Sigma_{\rm eff}$, $q=\sin^{2}\!\thetac$, and $\ln(\ell_{\mathrm{sc}}/a)=\tfrac12\ln(1/\Seff)$, so that the leading logarithmic law~\eqref{eq:tensionenergy} reads
\begin{equation}
W_{\mathrm{neck}}
=\pi\Seff\,q^{2}\ln\!\frac{\ell_{\mathrm{sc}}}{a}
=\frac{\pi}{2}\,\Seff\,\sin^{4}\!\thetac\,\ln\!\frac1\Seff .
\label{eq:verifpred}
\end{equation}
Subtracting the cap excess $\tau^{2}$ from~\eqref{eq:tensionexact} gives the complete analytic prediction for all $\thetac$ (see \S\ref{sub:tense}):
\begin{equation}
W_{\mathrm{full}}=\pi\Seff\Big\{\sin^{2}\!\thetac\,(1-\cos\thetac)+\sin^{4}\!\thetac\Big[\ln\frac{4\ell_{\mathrm{sc}}}{a}-\gamma_E-\tfrac12+\ln\frac{1-\cos\thetac}{\sin^{3}\thetac}\Big]\Big\}.
\label{eq:verifexact}
\end{equation}
Consequently, $W_{\mathrm{full}}/(\Seff\sin^{4}\!\thetac)$ is expected to grow linearly in $\ln(1/\Seff)$ with slope $\pi/2\simeq1.5708$ and the $\thetac$-dependent intercept of~\eqref{eq:verifexact}, and the charge to tend to $\sin^{2}\!\thetac$. Numerically, for $\Seff\in[10^{-4},10^{-1}]$ and $\thetac=45^\circ,60^\circ,75^\circ$, the slopes fitted at $\Seff\le2\times10^{-3}$ are $1.564$, $1.562$, $1.562$; $W_{\mathrm{full}}$ agrees with~\eqref{eq:verifexact} within $0.11\%$, $0.15\%$, $0.16\%$ for $\Seff\le10^{-3}$ (within $5.3\%$ overall). The charge read from $\rho\sin\psi/[xK_1(x)]$ at $\rho=\sqrt{\ell_{\mathrm{sc}} a}$ ($x=\rho/\ell_{\mathrm{sc}}$) converges to $\sin^{2}\!\thetac$ within $0.02\%$, $0.04\%$, $0.05\%$ at $\Seff=10^{-4}$.

\begin{figure}[t]
\centering
\includegraphics[width=0.86\textwidth]{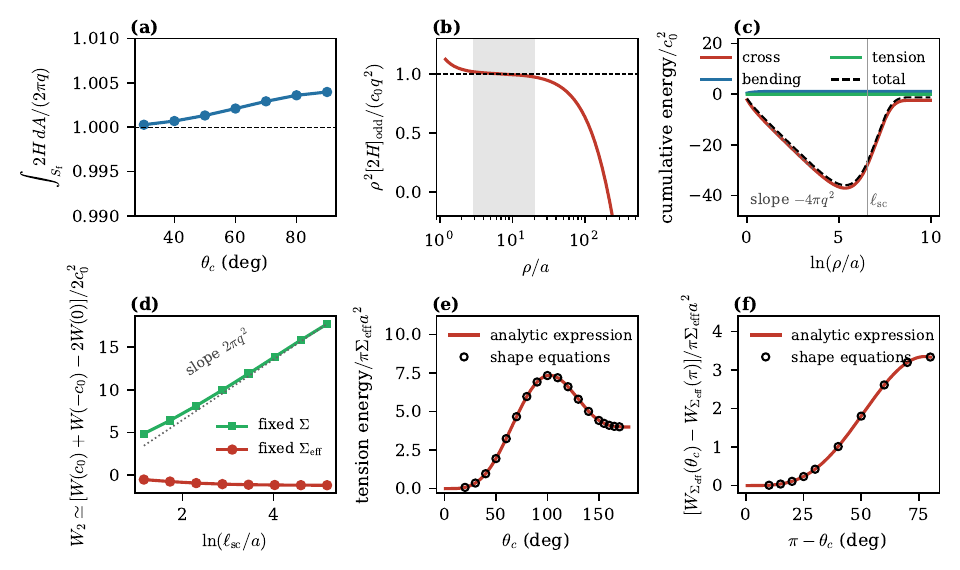}
\caption{Tests against the full nonlinear shape equations ($a=k=1$). \emph{Spontaneous curvature} ($\thetac=60^\circ$ except in (a)): (a)~$\int_{\mathcal S_{\mathrm f}}2H\,\dd A/(2\pi a\sin^{2}\!\thetac)$ at $\Seff=10^{-3}$, within $0.4\%$ of $1$. (b)~$\rho^{2}[2H]_{\rm odd}/(c_0q^{2})$ ($\Seff=2\times10^{-6}$, $c_0=\pm10^{-2}$): plateau at $1$ in the shaded overlap. (c)~Cumulative terms of~\eqref{eq:Eref} (parameters of (b)), even in $c_0$ and divided by $c_0^{2}$, against $\ln(\rho/a)$: the cross term (red) decreases with slope $-4\pi q^{2}$ (dotted) and is compensated at $\rho\sim\ell_{\mathrm{sc}}$ (vertical line); bending (blue) and tension (green) carry no logarithm; black dashed, total. (d)~Second central difference approximating $W_2$ against $\ln(\ell_{\mathrm{sc}}/a)$ ($c_0=\pm2\times10^{-3}$): it saturates at fixed $\Seff$ (red) and grows with slope $2\pi q^{2}$ (dotted) at fixed $\Sigma$ (green). \emph{Tension energy of the cap and of the neck} in units of $\pi\Seff a^{2}$ ($c_0=0$, $\Seff=10^{-4}$): (e)~against $\thetac$, the expression~\eqref{eq:tensionexact} (line) and the shape equations (circles; the cap term $\tau^{2}$ is added to the computed free-membrane energy). (f)~Near complete wrapping, $W_{\Seff}(\thetac)-W_{\Seff}(\pi)$ against $\pi-\thetac$: no term quadratic in $\chi$ is present.}
\label{fig:c0tests}
\end{figure}

\emph{Spontaneous-curvature tests} (Fig.~\ref{fig:c0tests}). At $c_0\neq0$ we perform four tests:

\emph{(i) Flux.} At $\Seff=10^{-3}$, $\int_{\mathcal S_{\mathrm f}}2H\,\dd A/(2\pi a\sin^{2}\!\thetac)$ stays within $0.4\%$ of unity across $30^\circ\le\thetac\le90^\circ$ (Fig.~\ref{fig:c0tests}a), confirming the cap--neck cancellation~\eqref{eq:totalHsum}.

\emph{(ii) Curvature tail.} At $\thetac=60^\circ$, $\Seff=2\times10^{-6}$, $c_0=\pm10^{-2}$, the odd part $\rho^{2}[2H]_{\rm odd}/(c_0q^{2})$ stays within $3\%$ of $1$ across the overlap $3\lesssim\rho/a\lesssim20$ before decaying into the outer field $qK_0/\ell_{\mathrm{sc}}^{2}$ (Fig.~\ref{fig:c0tests}b). The even bending energy has $c_0^{2}$ coefficient $2\pi\sin^{4}\!\thetac\,\tan^{2}(\thetac/2)$, matching the shape equations within $0.4\%$ at $\thetac=45^\circ,60^\circ,75^\circ$ ($\Seff=10^{-5}$).

\emph{(iii) Logarithm cancellation.} Accumulating the even parts of~\eqref{eq:Eref} up to radius $\rho$ (Fig.~\ref{fig:c0tests}c), the cross term falls with slope $-6.9\,c_0^{2}$ per unit $\ln\rho$ in $5\lesssim\rho/a\lesssim50$ (versus $-4\pi q^{2}c_0^{2}=-7.07\,c_0^{2}$ from~\eqref{eq:densities}), reaches $-37\,c_0^{2}$ near $\rho\simeq200$, and returns to $-2.4\,c_0^{2}$ beyond $\ell_{\mathrm{sc}}$ as required by~\eqref{eq:meanidentity}, while bending stays at $+1.2\,c_0^{2}$ and tension below $0.25\,c_0^{2}$. Consequently, the second central difference $[W_{\mathrm{full}}(c_0)+W_{\mathrm{full}}(-c_0)-2W_{\mathrm{full}}(0)]/(2c_0^{2})$, which approximates $W_2$ of~\eqref{eq:Eexpansion} for $k=1$, at fixed $\Seff$ saturates with slope $-0.03$ against $\ln(\ell_{\mathrm{sc}}/a)$ ($\thetac=60^\circ$, $c_0=\pm2\times10^{-3}$, Fig.~\ref{fig:c0tests}d). At fixed multiplier $\Sigma$, where $\Seff=\Sigma+2kc_0^{2}$ varies, it grows for $10^{-4}\le\Sigma\le10^{-3}$ (so that $2c_0^{2}\ll\Sigma$) with slope $3.41$ against $2\pi q^{2}=3.53$, that is, within $4\%$ of the slope due to the tension shift alone.

\emph{(iv) Composite solution.} Evaluating~\eqref{eq:Eref} directly on the parameter-free composite~\eqref{eq:composite}--{}\eqref{eq:alpha} matches $W_{\mathrm{full}}$ at $c_0=0$, $\Seff=10^{-4}$ within $0.02\%$ ($\thetac=30^\circ,60^\circ$) and $0.1\%$ ($85^\circ$; at $90^\circ$ the composite $\Psi_0=\arcsin(q/\rho)$ is singular at the contact line and the quadrature is not meaningful). At $\thetac=60^\circ$, $\Seff=10^{-5}$, $\partial_{c_0}W=-9.4252$ versus $-4\pi q=-9.4248$. The closed form $W_2=-2\pi a^{2}\tau^{3}(2-\tau)$ lies within $0.3\%$ of the shape equations at $45^\circ$, $60^\circ$, $75^\circ$. Evaluated on the composite solution, $W_2$ agrees with the shape equations within $1.5\%$ for $\thetac\le60^\circ$ ($-0.0283$ vs.\ $-0.0282$ at $30^\circ$; $-0.2690$ vs.\ $-0.2690$ at $45^\circ$; $-1.158$ vs.\ $-1.176$ at $60^\circ$, $\Seff=10^{-5}$) and by $7\%$ at $75^\circ$ ($-2.99$ vs.\ $-3.21$, due to the leading-slope approximation $q/\rho$ near the neck). Its variation per unit $\ln(\ell_{\mathrm{sc}}/a)$ at $60^\circ$ and $75^\circ$ does not exceed $0.006$ and $0.021$ in absolute value for $10^{-6}\le\Seff\le10^{-4}$, negligible compared to $2\pi q^{2}=3.5$ and $5.5$.

\emph{All-angle tension energy} (Fig.~\ref{fig:c0tests}e). At $c_0=0$ and $20^\circ\le\thetac\le170^\circ$ (including $\thetac>90^\circ$), $W_{\mathrm{full}}$ agrees with $\pi\Seff a^{2}[\mathcal A-\tau^{2}]$ from~\eqref{eq:tensionexact} within $0.16\%$ at $\Seff=10^{-3}$ and $0.03\%$ at $10^{-4}$, the difference vanishing as $\thetac\to\pi$ (Fig.~\ref{fig:c0tests}e), where no $\chi^{2}$ term appears, confirming~\eqref{eq:wrappedstab} (Fig.~\ref{fig:c0tests}f).

\let\oldthebibliography\thebibliography
\renewcommand\thebibliography[1]{\oldthebibliography{#1}\setlength{\itemsep}{0pt}\setlength{\parsep}{0pt}}
\bibliographystyle{unsrt}
\bibliography{BiblioBL}

\begin{thebibliography}{10}

\bibitem{Lipowsky1991}
R.~Lipowsky.
\newblock The conformation of membranes.
\newblock {\em Nature}, 349:475, 1991.

\bibitem{Seifert1997}
U.~Seifert.
\newblock Configurations of fluid membranes and vesicles.
\newblock {\em Adv. Phys.}, 46:13, 1997.

\bibitem{ACL2015}
J.~Agudo-Canalejo and R.~Lipowsky.
\newblock Critical particle sizes for the engulfment of nanoparticles by
  membranes and vesicles with bilayer asymmetry.
\newblock {\em ACS Nano}, 9:3704, 2015.

\bibitem{DesernoBickel2003}
M.~Deserno and T.~Bickel.
\newblock Wrapping of a spherical colloid by a fluid membrane.
\newblock {\em Europhys. Lett.}, 62:767, 2003.

\bibitem{Deserno2004}
M.~Deserno.
\newblock Elastic deformation of a fluid membrane upon colloid binding.
\newblock {\em Phys. Rev. E}, 69:031903, 2004.

\bibitem{NapoliGoriely2020}
G.~Napoli and A.~Goriely.
\newblock Elastocytosis.
\newblock {\em J. Mech. Phys. Solids}, 145:104133, 2020.

\bibitem{LipowskyDoebereiner1998}
R.~Lipowsky and H.-G. D\"obereiner.
\newblock Vesicles in contact with nanoparticles and colloids.
\newblock {\em Europhys. Lett.}, 43:219--225, 1998.

\bibitem{ACL2016}
J.~Agudo-Canalejo and R.~Lipowsky.
\newblock Stabilization of membrane necks by adhesive particles, substrate
  surfaces, and constriction forces.
\newblock {\em Soft Matter}, 12:8155, 2016.

\bibitem{ACL2017}
J.~Agudo-Canalejo and R.~Lipowsky.
\newblock Uniform and janus-like nanoparticles in contact with vesicles: energy
  landscapes and curvature-induced forces.
\newblock {\em Soft Matter}, 13:2155, 2017.

\bibitem{BiscariNapoli2007}
P.~Biscari and G.~Napoli.
\newblock Inclusion-induced boundary layers in lipid vesicles.
\newblock {\em Biomech. Model. Mechanobiol.}, 6:297, 2007.

\bibitem{Xiao2022}
K.~Xiao, R.~Ma, and C.-X. Wu.
\newblock Force-induced wrapping phase transition in activated cellular uptake.
\newblock {\em Phys. Rev. E}, 106:044411, 2022.

\bibitem{Canham1970}
P.~B. Canham.
\newblock The minimum energy of bending as a possible explanation of the
  biconcave shape of the human red blood cell.
\newblock {\em J. Theor. Biol.}, 26:61, 1970.

\bibitem{Helfrich1973}
W.~Helfrich.
\newblock Elastic properties of lipid bilayers: theory and possible
  experiments.
\newblock {\em Z. Naturforsch. C}, 28:693, 1973.

\bibitem{Necks2025}
F.~Fessler, P.~Muller, and A.~Stocco.
\newblock Energetics and dynamics of membrane necks in particle wrapping.
\newblock {\em J. Colloid Interface Sci.}, 700:138524, 2025.

\bibitem{SeifertBerndlLipowsky1991}
U.~Seifert, K.~Berndl, and R.~Lipowsky.
\newblock Shape transformations of vesicles: Phase diagram for
  spontaneous-curvature and bilayer-coupling models.
\newblock {\em Phys. Rev. A}, 44:1182, 1991.

\bibitem{capovilla2002stresses}
R.~Capovilla and J.~Guven.
\newblock Stresses in lipid membranes.
\newblock {\em J. Phys. A: Math. Gen.}, 35:6233, 2002.

\bibitem{hinch1991perturbation}
E.~J. Hinch.
\newblock {\em Perturbation Methods}.
\newblock Cambridge Texts in Applied Mathematics. Cambridge University Press,
  Cambridge, 1991.

\bibitem{osserman1986survey}
R.~Osserman.
\newblock {\em A Survey of Minimal Surfaces}.
\newblock Dover Publications, New York, 2nd edition, 1986.

\bibitem{muller2005geometry}
M.~M. M\"uller, M.~Deserno, and J.~Guven.
\newblock Geometry of surface-mediated interactions.
\newblock {\em Europhys. Lett.}, 69:482, 2005.

\bibitem{deserno2015fluid}
M.~Deserno.
\newblock Fluid lipid membranes: From differential geometry to curvature
  stresses.
\newblock {\em Chem. Phys. Lipids}, 185:11, 2015.

\bibitem{abramowitz1972handbook}
M.~Abramowitz and I.~A. Stegun.
\newblock {\em Handbook of Mathematical Functions with Formulas, Graphs, and
  Mathematical Tables}.
\newblock Dover Publications, New York, 1972.

\bibitem{Boulant2011}
S.~Boulant, C.~Kural, J.-C. Zeeh, F.~Ubelmann, and T.~Kirchhausen.
\newblock Actin dynamics counteract membrane tension during clathrin-mediated
  endocytosis.
\newblock {\em Nat. Cell Biol.}, 13:1124, 2011.

\bibitem{MechanoCME}
J.~P. Ferguson, S.~D. Huber, N.~M. Willy, E.~Ayg\"un, S.~Goker, T.~Atabey, and
  C.~Kural.
\newblock Mechanoregulation of clathrin-mediated endocytosis.
\newblock {\em J. Cell Sci.}, 130:3631, 2017.

\bibitem{Dietrich1997}
C.~Dietrich, M.~Angelova, and B.~Pouligny.
\newblock Adhesion of latex spheres to giant phospholipid vesicles.
\newblock {\em J. Phys. II France}, 7:1651, 1997.

\bibitem{Spanke2020}
H.~T. Spanke, R.~W. Style, C.~Fran\c{c}ois-Martin, M.~Feofilova, M.~Eisentraut,
  H.~Kress, J.~Agudo-Canalejo, and E.~R. Dufresne.
\newblock Wrapping of microparticles by floppy lipid vesicles.
\newblock {\em Phys. Rev. Lett.}, 125:198102, 2020.

\bibitem{OptTweez2023}
F.~Fessler, V.~Sharma, P.~Muller, and A.~Stocco.
\newblock Entry of microparticles into giant lipid vesicles by optical
  tweezers.
\newblock {\em Phys. Rev. E}, 107:L052601, 2023.

\bibitem{Ayala2023}
Y.~A. Ayala, R.~Omidvar, W.~R\"omer, and A.~Rohrbach.
\newblock Thermal fluctuations of the lipid membrane determine particle uptake
  into giant unilamellar vesicles.
\newblock {\em Nat. Commun.}, 14:65, 2023.

\bibitem{Strobl2014}
F.~G. Strobl, F.~Seitz, C.~Westerhausen, A.~Reller, A.~A. Torrano,
  C.~Br\"auchle, A.~Wixforth, and M.~F. Schneider.
\newblock Intake of silica nanoparticles by giant lipid vesicles: influence of
  particle size and thermodynamic membrane state.
\newblock {\em Beilstein J. Nanotechnol.}, 5:2468, 2014.

\bibitem{Reynwar2007}
B.~J. Reynwar, G.~Illya, V.~A. Harmandaris, M.~M. M\"uller, K.~Kremer, and
  M.~Deserno.
\newblock Aggregation and vesiculation of membrane proteins by
  curvature-mediated interactions.
\newblock {\em Nature}, 447:461, 2007.

\bibitem{MidyaAuthGompper2023}
J.~Midya, T.~Auth, and G.~Gompper.
\newblock Membrane-mediated interactions between nonspherical elastic
  particles.
\newblock {\em ACS Nano}, 17:1935, 2023.

\bibitem{JulicherLipowsky1993}
F.~J\"ulicher and R.~Lipowsky.
\newblock Domain-induced budding of vesicles.
\newblock {\em Phys. Rev. Lett.}, 70:2964, 1993.

\bibitem{Li2017Langmuir}
N.~Li, N.~Sharifi-Mood, F.~Tu, D.~Lee, R.~Radhakrishnan, T.~Baumgart, and K.~J.
  Stebe.
\newblock Curvature-driven migration of colloids on tense lipid bilayers.
\newblock {\em Langmuir}, 33:600, 2017.

\bibitem{KraljIglic1999}
V.~Kralj-Igli\v{c}, V.~Heinrich, S.~Svetina, and B.~\v{Z}ek\v{s}.
\newblock Free energy of closed membrane with anisotropic inclusions.
\newblock {\em Eur. Phys. J. B}, 10:5, 1999.

\bibitem{kierzenka2001bvp}
J.~Kierzenka and L.~F. Shampine.
\newblock A bvp solver based on residual control and the {MATLAB} {PSE}.
\newblock {\em ACM Trans. Math. Software}, 27:299, 2001.

\end{thebibliography}

\end{document}